\documentclass[letter,10pt]{iopart}

\usepackage{amsfonts}
\usepackage{latexsym}
\usepackage{eucal}
\usepackage{graphicx}
\usepackage{psfrag}
\usepackage{color}
\usepackage[final]{changes}
\usepackage{lineno}
\usepackage{array}
\newcolumntype{L}[1]{>{\raggedright\arraybackslash}p{#1}}
\usepackage{booktabs}
\usepackage{adjustbox}
\usepackage{rotating}
\usepackage{hyperref}

\def\coloneqq{\mathrel{\vcenter{\baselineskip0.5ex \lineskiplimit0pt
                     \hbox{\scriptsize.}\hbox{\scriptsize.}}}%
                     =}

\def\dd{\mbox{d}}

\def\ve{\varepsilon}

\def\n{{\bf n}}
\def\m{{\bf m}}

\def\x{{\bf x}}

\def\S{{\bf S}}

\def\cb{\color{blue}}

\usepackage{dsfont}
\usepackage{enumerate}

\usepackage{iopams}  

\begin{document}

\title[Biological Diversity]{Diversity in Biology: definitions,
  quantification, and models}

\author{Song Xu$^{1}$, Lucas B\"{o}ttcher$^{2}$, and Tom Chou$^{3}$}

\address{$^{1}$Center for Biomedical Informatics Research, Department
  of Medicine, Stanford University, Stanford, CA, 94305, USA}
\address{$^{2}$Institute for Theoretical Physics and Center of
  Economic Research, ETH Zurich, Zurich, Switzerland}
\address{$^{3}$Department of Mathematics, UCLA, Los Angeles, CA,
  90095-1766, USA}

\ead{tomchou@ucla.edu}


\begin{abstract}
Diversity indices are useful single-number metrics for characterizing
a complex distribution of a set of attributes across a population of
interest. The utility of these different metrics or sets of metrics
depend on the context and application, and whether a predictive
mechanistic model exists. In this topical review, we first summarize
the relevant mathematical principles underlying heterogeneity in a
large population before outlining the various definitions of
`diversity' and providing examples of scientific topics in which its
quantification plays an important role. We then review how diversity
has been a ubiquitous concept across multiple fields including
ecology, immunology, cellular barcoding
\replaced{experiments}{studies}, and socioeconomic studies. Since many
of these applications involve sampling of populations, we also review
how diversity in small samples is related to the diversity in the
entire population. Features that arise in each of these applications
are highlighted.
\end{abstract}

%

\submitto{Physical Biology}
%
%
%
\ioptwocol
\begin{figure*}[t!]
\centering
\includegraphics[width=5.1in]{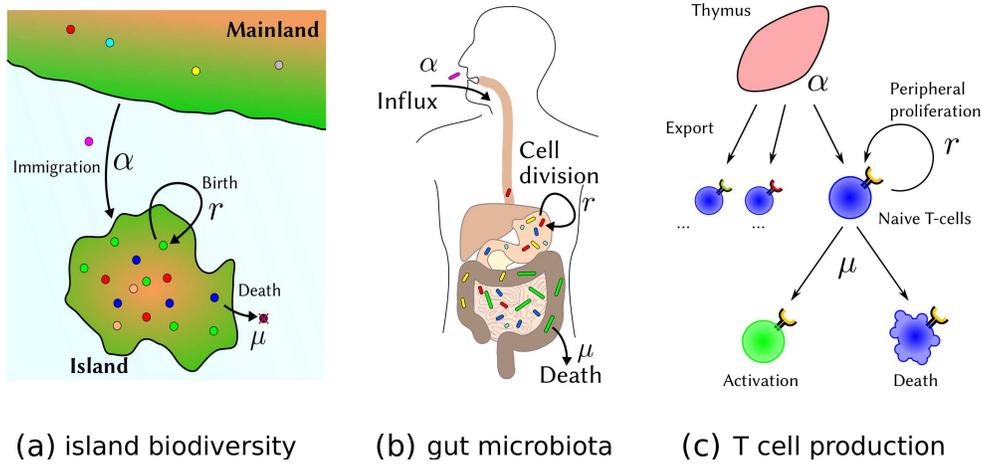}
\caption{Examples of complex, multicomponent populations in
    which diversity may be a meaningful quantitative concept. (a)
  Diversity in island ecology. A large number of species may migrate
  onto an island. Organisms can proliferate and die, leading to a
  specific time-dependent pattern of species diversity on the
  island. (b) Microbes are ingested and form a community in the gut by
  proliferating, competing, and dying.  They can also be cleared from
  the gut. (c) Naive T cell generation in vertebrates. Naive T cells
  develop in the thymus. Each T cell expresses only one type of T cell
  receptor (TCR). Naive T cells can proliferate and die in the
  peripheral blood. The possible number of T cell receptors that can
  be expressed is enormous $> 10^{15}$, but only perhaps $10^6 - 10^8$
  different TCRs usually exist in an organism. The diversity of the T
  cell receptor repertoire is an important determinant of the
  organism's response to antigens.
\label{FIG1}}
\end{figure*}
\section{Introduction}
Diversity is a frequently used concept across a broad spectrum of
scientific disciplines, ranging from
biology~\cite{nei1973analysis,heywood1995global,purvis2000getting,WHITTAKER2001,sala2000global}
and
ecology~\cite{fisher1943relation,magurran1988ecological,benton1995diversification,courtillot1996effects,alroy2004sepkoski,stollmeier2014possible},
\replaced{to}{over} investment and portfolio
theory~\cite{blume1974stock,blume1975asset,rajan2000cost,goetzmann2008equity,haldane2011systemic},
to linguistics~\cite{greenberg1956measurement,yule2014statistical} and
sociology~\cite{bailey1990social,balch2000hierarchic,neckerman2007inequality,ostrom2009understanding,mas2013short,domina2017categorical}.
In each of these disciplines, diversity is a measure of the range and
distribution of certain features within a given population. It is
considered a key attribute that can be dynamically varying, influenced
by intra-population interactions, and modified by environmental
factors. The concept of diversity, variety, or heterogeneity can be
applied to any population. The evolution of the population can also be
highly correlated with its diversity. Some examples of biological
population dynamics occurring at different scales are shown in
Fig.~\ref{FIG1}. At first sight, diversity seems to be an intuitively
simple concept, but since certain population attributes require a full
distribution function to quantify, it can be rather complex and
difficult to capture using a single metric
\cite{sepkoski1988alpha,purvis2000getting,WHITTAKER2001,ricotta2005through}. We
could for example think of a community with a total of four species,
with one of the species dominating the total population.  Consider a
second community that consists of two equally common species. Which
one of the two communities exhibits a higher diversity?  The first
one, because it harbors a larger number of species? Or the second one,
because a sample is more likely to contain two species?  This example
shows that diversity is intrinsically linked to the total number of
extant species (richness) and how the population is distributed
throughout the species (evenness), and thus cannot be captured by a
single number~\cite{purvis2000getting}. As a result, there are
numerous different diversity indices and associated concepts used in
different
applications~\cite{simpson1949measurement,hurlbert1971nonconcept,sepkoski1988alpha,purvis2000getting,WHITTAKER2001,ricotta2005through,sarkar2006ecological}.
Nonetheless, diversity measures are important for assessing the
current condition of ecosystems, \replaced{quantifying}{to quantify}
the influence of environmental factors on different species, and
\replaced{planning conservation efforts}{in the context of
  conservation
  planning}~\cite{heywood1995global,courtillot1996effects,john1998rates,sala2000global,margules2000systematic,alroy2004sepkoski,sarkar2006ecological}.
In addition, the concept of diversity is important for the
quantitative description of wealth distributions and, more generally,
\replaced{for}{to} identifying mechanisms leading to variations in
societies~\cite{BELLCURVE,LORENZ1905,HOOVER1936,ESTEBAN_RAY1994,DUCLOS2004}.
In a broader sense, diversity indices may be helpful for the design of
robust energy distribution systems~\cite{grubb2006diversity} or even
to assemble well-performing teams~\cite{mas2013short}. Thus we see
that, despite the ambiguity in the definition of diversity, the
concept is very relevant to many different disciplines and
applications.

In this topical review, we start by summarizing the basic concepts
from information theory which are necessary for a quantitative
treatment of diversity. We continue with describing aspects of
populations and diversity that are common to many applications in
biology. In the next section, we present the common mathematical
descriptions of diversity in terms of both number and species
counts. Moreover, in most applications, only a small sample of a
population is available. Thus, we place particular emphasis on the
effects of sampling on diversity measures in
Section~\ref{sec:SAMPLING}. In Section \ref{sec:FIELDS} and
subsections within, we survey a number of biological systems in which
concepts of diversity play a key role in understanding the dynamics of
the population. These include ecological populations, stem cell
barcoding experiments, immunology, cancer, and societal wealth
distributions. Each of these systems carry their unique attributes and
thus require specific diversity measures. Finally, in
Section~\ref{sec:summary} we summarize the advantages and
disadvantages of some common diversity measures and conclude with a
discussion of possible future applications of concepts of diversity.
%
%
%


\section{Entropy, relative entropy, 
KL divergence, KS statistic, mutual information and all that}

%
%
%

We first provide a summary of the fundamental mathematical structures
that arise in the analysis of populations in which one naturally seeks
to quantitatively compare distributions or frequencies of
subpopulations. These mathematical notions invariably involve ideas
from information theory such as entropy and mutual information which
have a rich history and deep connections to thermodynamics, coding
theory, cryptography, inference, and
communication~\cite{cover2012elements}. To review the necessary
information-theoretic concepts, we consider a discrete random variable
$X$ which takes on values from the set $\{x_1,x_2,\dots,x_N\}$ with
probability $P_k=\mathrm{Pr}\left(X=x_k\right)$ such that
\begin{equation}
\sum_{k=1}^{N} P_k = 1,
\end{equation}
where the sum is taken over all possible values $x_{k}$. This
probability mass function may represent the relative frequency that
\replaced{an}{the} attribute $X$ takes on the value $x_{k}$
\replaced{within}{in} a large population.  In the case of species
diversity, we may interpret $P_k$ as the \added{relative frequency of
  species $k$} or \replaced{the fraction of species with trait $X=x_k$
  (see clone counts in Section \ref{CLONECOUNTS}).}{in a certain
  population.}  

The entropy, or `Shannon entropy\added{,}' is defined by
\begin{equation}
H(X) = -\sum_{k=1}^{N}P_k\log P_k.
\label{ENTROPY}
\end{equation}
and can be thought of as the expected uncertainty or surprise
$-\mathds{E}[\log P(X)]$.

The continuous limit of Shannon entropy, or \textit{differential}
Shannon entropy\added{,} has also been defined, but care must be taken if $X$
carries physical dimensions. If the probability of $X$ taking on
values in the interval $[x, x+\dd x]$ is denoted by $P(x)\, \dd x$,
the differential Shannon entropy is
\begin{equation}
H(X) = - \int
P(x)\log P(x)\, \dd x\,.
\label{eq:h_cont}
\end{equation}
These expressions are synonymous with the `Shannon index' of species
diversity with some freedom in the choice of the base of the
logarithm. Without any constraints on the distributions other than
being compactly supported, the form of $P_{k}$ or $P(x)$ that
maximizes $H(X)$ is a uniform distribution. With additional
constraints there are classes of distributions that maximize the
Shannon index. For example, for a fixed mean and variance on an
unbounded domain, the Shannon index- or entropy-maximizing
distribution is Gaussian. Within Gaussian distributions, the Shannon
index increases logarithmically with the variance. In fact, within a
specific class of distributions, the Shannon index is larger for
flatter distributions \cite{JAYNES1963,LAZO}.  As such, the Shannon
index has been used as a measure of
diversity~\cite{spellerberg2003tribute}.
  
\replaced{One issue}{The problem} with the differential entropy of Eq.~(\ref{eq:h_cont}) is
that $P(x)$ carries dimensions $X^{-1}$, because the cumulative
distribution function $P(X\leq x)=\int_{-\infty}^{x} P(x') \, \dd x'$
has to be dimensionless. Therefore, the argument of the logarithm in
Eq.~(\ref{eq:h_cont}) is not dimensionless as required. To avoid such
an issue, one can define a point-density function $P_0(x)$ according
to~\cite{JAYNES1963}
\begin{equation}
\lim_{N\to \infty}{\# {\rm points} \in [a,b]\over N} \equiv  \int_{a}^{b} P_{0}(x)\dd x.
\end{equation}
Given that the limit is well-behaved, we can express the difference
between two adjacent points $x_{k+1}$ and $x_k$ in terms of
\begin{equation}
\lim_{N\to \infty}\left[N (x_{k+1}-x_{k})\right]= P_{0}^{-1}(x_k).
\end{equation}
We now consider the continuum limit of the discrete Shannon entropy as
defined in Eq.~(\ref{ENTROPY}), and set
\begin{equation}
P_k=P(x_k)(x_{k+1}-x_k)=P(x_k) \left[N P_0(x_k)\right]^{-1}.
\end{equation}
In this way, it is possible to derive a continuous Shannon entropy
\begin{eqnarray}
\lim_{N\to \infty}H_{N}(X) &= -\int P(x)\log\left({P(x) \over N
  P_{0}(x)}\right)\dd x-\log(N)\nonumber \\ &=-\int
P(x)\log\left({P(x) \over P_{0}(x)}\right)\dd x
\label{LIMITINGDENSITY}
\end{eqnarray}
that is invariant under parameter changes and whose logarithm depends
on the dimensionless quantity $P(x)/ P_{0}(x)$. We subtracted
$\log(N)$ in Eq.~(\ref{LIMITINGDENSITY}) to obtain a finite
$H_{N}(X)$.

To characterize the diversity between two communities, we consider two
discrete random variables $X$ and $Y$ with the corresponding joint
probability mass function
$P_{X,Y}(x_k,y_{\ell})=\mathrm{Pr}(X=x_k,Y=y_{\ell})$. Given the joint
distribution $P_{X,Y}(x_k,y_{\ell})$, we can compute the marginal
distributions $P_X(x_k)=\sum_{\ell} P_{X,Y}(x_k,y_{\ell})$ and
$P_Y(y_{\ell})=\sum_{k} P_{X,Y}(x_k,y_{\ell})$ by summing over the
complementary variable. These definitions enable us to define the
joint entropy
\begin{equation}
H(X,Y) = -\sum_{k,\ell}P_{X,Y}(x_{k}, y_{\ell})\log
P_{X,Y}(x_{k}, y_{\ell}),
\label{JOINTENTROPY}
\end{equation}
which may be also written as $-\mathds{E}[\log P_{X,Y}]$. Moreover,
the conditional entropy
\begin{eqnarray}
H(Y\vert X) & = -\sum_{k,\ell}P_{X,Y}(x_{k}, y_{\ell})\log
\left({P_{X,Y}(x_{k},y_{\ell})\over P_{X}({x}_{k})}\right) \nonumber \\
\: & = -\sum_{k,\ell}P_{X,Y}(x_{k}, y_{\ell})\log
P_{Y \vert X}(y_{\ell}\vert x_{k})
\label{CONDENTROPY}
\end{eqnarray}
describes the expected uncertainty in the random variable $Y$ given
$X$. It can be also expressed as $-\mathds{E}[\log P_{Y\vert X}]$
where $P_{Y\vert X}$ is the conditional probability mass function.
From symmetry, Eq.~(\ref{CONDENTROPY}) also holds when all $X$ and $Y$
are interchanged. For independent random variables $X$ and $Y$, we
find that $H(Y\vert X) = H(Y)$ and $H(X\vert Y) = H(X)$.

While the Shannon index is a measure of the absolute entropy of a
distribution, the relative entropy or Kullback-Leibler (KL) divergence
\begin{eqnarray}
D_{\mathrm{KL}}(P\| Q) & = \sum_{k}P(x_{k})\log\left({P(x_{k})\over Q(x_{k})}\right) \nonumber \\
\: & = \mathds{E}_{P}\left[\log P(x_{k}) - \log Q(x_{k})\right],
\label{KLDIVERGENCE}
\end{eqnarray}
quantifies the distance between two probability mass functions $P$ and
$Q$. In the case of continuous distributions $P(x)$ and $Q(x)$, we
obtain $D_{\mathrm{KL}}(P\| Q) = \int P(x)\log(P(x)/Q(x))\, \dd x$.

The KL divergence is the relative entropy of $P$ with respect to the
reference distribution $Q$. Note that the limiting Shannon entropy is
simply the KL divergence between the distribution $P(x)$ and the
associated invariant measure $P_{0}(x)$. Usually, $P$ is an
experimental or observed distribution and $Q$ is a model that
represents $P$. Furthermore, the KL divergence is nonnegative and
equals zero if and only if $P=Q$~\cite{cover2012elements}. It is not
symmetric, $D_{\mathrm{KL}}(P\| Q) \neq D_{\mathrm{KL}}(Q\| P)$, and
is thus not a metric. In addition, a special case of the KL divergence
is the `mutual information'
\begin{eqnarray}
I(X;Y) & = D_{\mathrm{KL}}(P_{X,Y} \| P_{X}P_{Y}) \nonumber \\
\: & = \sum_{k, \ell} P_{X,Y}(x_{k},y_{\ell})\log\left({P_{X,Y}(x_{k},y_{\ell})\over
P_{X}(x_{k})P_{Y}(y_{\ell})}\right)\!.
\label{I}
\end{eqnarray}
%
%
Note that $I(X;Y)= I(Y;X)$ is symmetric and quantifies how much
knowing one variable reduces the uncertainty in the other. If $X$ and
$Y$ are completely independent, $I(X,Y) = 0$. According to
Eq.~(\ref{I}) and the definitions of joint and conditional entropy in
Eqs.~(\ref{CONDENTROPY}) and (\ref{JOINTENTROPY}), the mutual
information can be written in terms of marginal, conditional, and
joint entropies~\cite{cover2012elements}:
\begin{eqnarray}
I(X;Y) & = H(X)-H(X\vert Y) = H(Y)-H(Y\vert X) \nonumber \\
\: & = H(X)+H(Y)-H(X,Y).
\end{eqnarray}
A symmetric version of the KL divergence is provided by the
Jensen-Shannon divergence~\cite{lin1991divergence}
\begin{equation}
{\rm JSD}(P\|Q) = {1\over 2}D_{\mathrm{KL}}(P\|M)+{1\over 2}D_{\mathrm{KL}}(Q\|M),
\end{equation}
where $M = (P+Q)/2$ defines the mean distribution of $P$ and $Q$.
These divergences can be extended to include multiple and
higher-dimensional distributions. The square-root of the
Jensen-Shannon divergence is a distance metric between two
distributions.

Another useful distance metric is the Kolmogorov-Smirnov (KS)
distance, which is defined as
\begin{equation}
D_{\mathrm{KS}} = \max_{x} \vert G(x)-F(x)\vert,
\label{KSMETRIC}
\end{equation}
where $F(x)$ is a cumulative reference distribution and $G(x)$ is an
empirical distribution function. The distribution $G(x)$ is based on
different samples with cumulative distribution function that can be
$F(x)$ or another distribution to be tested against $F(x)$.  The KS
metric is the maximum distance between the two cumulative
distributions $F(x)$ and $G(x)$. We outline in
Section~\ref{sec:societal} that the KS metric is related to the Hoover
index which is used to quantify diversity, or inequity, in wealth or
income distributions relative to a uniform distribution.
\section{Commonly used measures of diversity}
The notions of entropy and information are naturally related to the
spread of a distribution $P(x)$, and can be subsumed into a general
metric for quantifying diversity. Usually, a population is measured
and can be thought of as one realization of an underlying
distribution. Consider a realization $\n = \{n_{1}, n_{2}, \ldots,
n_{R}\}$ describing the number $n_{i}$ of entities of a discrete and
distinguishable group/species/type ($1\leq i\leq R$). The total
population is $N = \sum_{i=1}^{R}n_{i}$. This given realization
constitutes a `distribution' across all possible types. Thus, any
realization is completely described by a set of $R$ numbers.
%
%
Diversity measures are reduced representations of the distribution. An
example would be a single parameter which captures the spread of the
distribution of realizations $\{n_{i}\}$. This is not different than,
for example, defining a Gaussian distribution by its mean and standard
deviation. Realizations $\{n_{i}\}$, however, usually are not
described by specific functions that can be defined by one or two
parameters such as Gaussians. However, many different diversity
indices can be unified into a single formula called `Hill numbers'
of order $q$~\cite{HILL1973,TUOMISTO2010A,TUOMISTO2010B}:
\begin{equation}
^{q}\!D = \left(\sum_{i=1}^{R} f_{i}^{q}\right)^{1/(1-q)},
\label{DEQN}
\end{equation}
where $f_{i} \equiv n_{i}/N$ is the relative abundance of types
$i$. This general formula represents different classes of `diversity
indices' for different values of $q$.  It is also useful because one
can consistently define an \textit{effective} proportional abundance
\begin{equation}
f_{\rm eff}\coloneqq 1/^{q}\!D =  \left(\sum_{i=1}^{R} f_{i}^{q}\right)^{1/(q-1)}
\end{equation}
that corresponds to an average abundance with increasing weighting
towards the larger-population species as $q$ increases
\cite{jost2007partitioning,TUOMISTO2010B}.

Note the similarity of this definition to the standard mathematical
$p$-norm
\begin{equation}
\vert\vert{\bf f}\vert\vert_{p} \coloneqq \left(\sum_{i=1}^{R} f_{i}^{p}\right)^{1/p},
\label{DEQNF}
\end{equation}
except that the exponent is $1/p$ instead of $1/(1-q)$. Another
diversity measure is provided by the Renyi
index~\cite{renyi1961measures}
\begin{equation}
^{q}\!H = \log {}^{q}\!D = {1\over 1-q}\log \left(\sum_{i=1}^{R} f_{i}^{q}\right),
\end{equation}
which is a generalization of the Shannon entropy defined in
Eq.~(\ref{ENTROPY}). The order $q$ describes the sensitivity of
$^{q}\!D$ and $^{q}\!H$ to common and rare
types~\cite{jost2006entropy}. Below, we provide an overview of the
most commonly used indices which result from the generalized diversity
$^{q}\!D$ for different values of $q$:


\vspace{2mm}

\emph{Richness.}---In the limit of $q\to 0^{+}$, the probabilities
$f_{i}^q$ are equal to unity and $^{0}\!D$ is simply the total number
of types in the population, or the `richness' $R$.  The richness is
often used in quantifying the diversity of T cells and species counts
in ecology~\cite{purvis2000getting} and represents a metric that
weights \replaced{all subpopulations equally}{the smallest subpopulations the most}.

\vspace{2mm}

\emph{Shannon index.}---For $q = 1-\varepsilon$ in the limit
$\epsilon\to 0^+$, the generalized diversity as defined by
Eq.~(\ref{DEQN}) becomes
\begin{eqnarray}
^{1}\!D & = \!\lim_{\ve \to 0^{+}}\left(\sum_{i=1}^{R}f_{i}^{1-\ve}\right)^{1/\ve} \!\!\!\! =
\lim_{\ve \to 0^{+}}\left(\sum_{i=1}^{R} f_{i} e^{-\ve \ln f_{i}}\right)^{1/\ve} \nonumber \\
\: & = \lim_{\ve \to 0^{+}} \left(\sum_{i=1}^{R} f_{i}(1-\ve \ln f_{i} 
+ O(\ve^{2}))\right)^{1/\ve}\nonumber  \\
\: & = \lim_{\ve \to 0^{+}}\left(1-\ve \sum_{i=1}^{R}f_{i}\ln f_{i}\right)^{1/\ve} \nonumber \\
\: & = \exp\left[-\sum_{i=1}^{R}f_{i}\ln f_{i}\right],
\end{eqnarray}
which is the exponential of the Shannon index

\begin{equation}
{\rm Sh}\coloneqq \ln\left(\lim_{q\to 1^{-}} {}^{q}\!D\right) =
-\sum_{i=1}^{R}f_{i}\log f_{i}
\label{SHANNONDIV}
\end{equation}
that parallels the Shannon entropy defined in Eqs.~(\ref{ENTROPY}) and
(\ref{CONDENTROPY}). This index is also sometimes called the
Shannon-Wiener index ($H$) and can be defined using any logarithmic
base. Usually measured values are ${\rm Sh} \sim {\cal
  O}(1)$. Qualitatively, $e^{\rm Sh}$ can be thought of as a rule of
thumb for the number of effective species in a population.

\vspace{2mm}

\emph{Evenness.}---Evenness is another class of diversity indices
often invoked in ecological and sociological studies. One definition
(`Shannon's equitability') is based on simply normalizing the Shannon
diversity by the maximum Shannon diversity that arises if every
species is equally likely \cite{EVENNESS1974}:
\begin{equation}
J_{\rm E} \coloneqq {{\rm Sh} \over {\rm Sh}_{\rm max}} = {{\rm Sh} \over \ln R}.
\end{equation}

\vspace{2mm}

\emph{Simpson's index with replacement.}---
When $q=2$, we find
\begin{equation}
^{2}\!D = 1/\left(\sum_{i=1}^{R} f_{i}^{2}\right).
\label{SIMPSONINDEX0}
\end{equation}
Simpson's diversity index is defined as 
\begin{equation}
S_{\rm r} = 1/{^{2}\!D} = \sum_{i=1}^{R} f_{i}^{2} = \sum_{i=1}^{R}
\left({n_{i}\over N}\right)^{2},
\label{SIMPSONINDEX}
\end{equation}
which carries the interpretation that upon drawing an entity 
from a given population the same type is selected twice.

\vspace{2mm}

\emph{Simpson's index without replacement.}---A related index that
cannot be directly constructed from $^{q}\!D$ is Simpson's index
\emph{without} replacement:
\begin{equation}
S = \sum_{i=1}^{R} {n_{i}(n_{i}-1)\over N (N-1)}.
\label{SIMPSONINDEX2}
\end{equation}
Here, when an entity is drawn, it is not replaced before the second
entity is drawn. The differences between $S_{\rm r}$ and $S$ are
significant only for systems with small numbers of entities $n_{i}$
for all types $i$.

\vspace{2mm}

\emph{Berger-Parker diversity index.}---In the $q \to \infty$ limit, we find
\begin{eqnarray}
^{\infty}\!D &=\lim_{q\to \infty}\left(\sum_{i=1}^{R} f_{i}^{q}\right)^{1/(1-q)} \nonumber \\
&=\lim_{q\to \infty} f_{\rm max}^{-\frac{1}{1-1/q}}\left[\sum_{i=1}^{R} 
\left(\frac{f_{i}}{f_{\rm max}}\right)^{q}\right]^{1/(1-q)} \nonumber \\
&= f_{\rm max}^{-1}
\end{eqnarray}
where $f_{\rm max}=\max_{i\in\{1,\dots,R\}}(f_i)$. The Berger-Parker diversity
  index 
\begin{equation}
1/{^{\infty}\!D}\coloneqq f_{\rm max}
\label{BERGERPARKER}
\end{equation}
is defined as the maximum abundance in the set $\{f_i\}$,
\textit{i.e.}, the abundance of the most common species.  It is
equivalent to the optimal solution of an $\infty$-norm of ${\bf
  f}={\bf n}/N$.

\begin{figure*}[htb]
\centering
\hspace{-1mm}\includegraphics[width=6.4in]{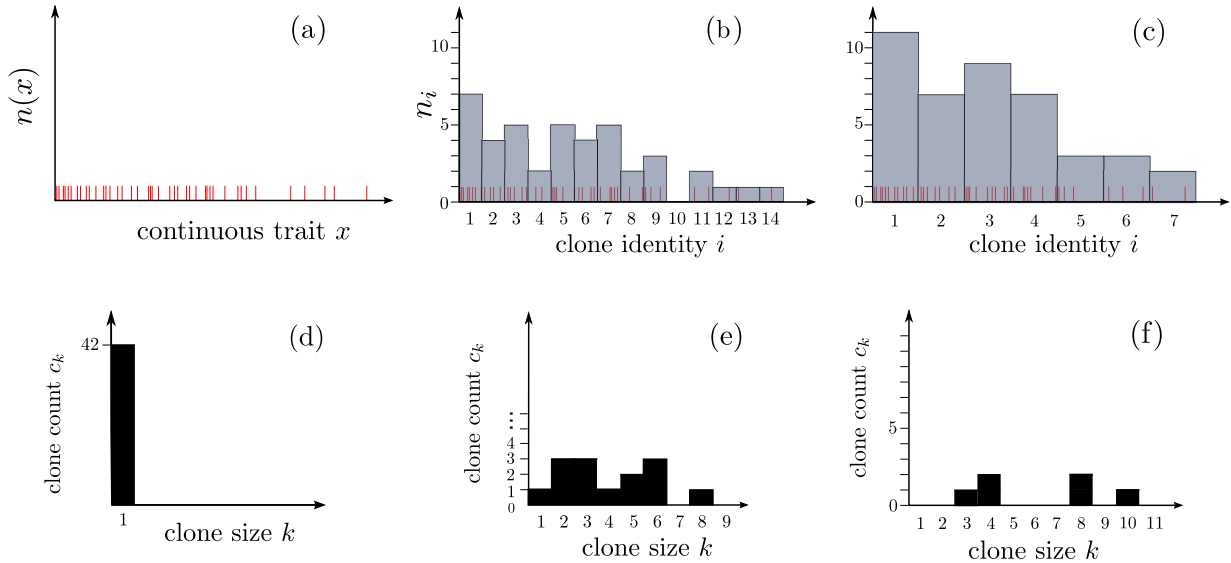}
\caption{Number counts and clone counts vary depending on the
  definition and binning of \replaced{traits or species
    identity}{discrete species}. \added{Both the number counts and
    clone count distributions can vary significantly as the
    distinguishability threshold is changed as shown in (a)-(c) and
    (d)-(f) where the resolution is coarsened.} \deleted{This
    consideration arises in designing experimental measurements.}}
\label{CONFIGS}
\end{figure*}

\section{Clone count representation}
\label{CLONECOUNTS}
An alternative way of quantifying a population is through the 
species abundance distribution or `clone counts' defined by 
\begin{equation}
c_{k} \coloneqq \sum_{i=1}^R\mathds{1}(n_{i},k) \in \mathbb{Z}^{+},
\label{CKEQN}
\end{equation}
where the discrete indicator function $\mathds{1}(n,k)=1$ if $n=k$ and
zero otherwise. The sum is usually taken over all species for which
$n_{i} \geq 1$.  Clone counts can also be defined over only a certain
special subset of species. Clone counts, or species abundance
distributions\deleted{,} in the language of computational mathematics, can be
thought of as the measure of the \textit{level-sets} \cite{SETHIAN} of
the discrete function $n_{i}$, or, in the language of condensed matter
physics, the \textit{density of states} if $n_{i}$ are thought of as
energies of states $i$ \cite{DENSITYOFSTATES}. The clone counts also
satisfy
\begin{equation}
N = \sum_{k=1}^{\infty}kc_{k} \quad \mbox{and}\quad  R= \sum_{k=1}^{\infty}c_{k},
\end{equation}
where $N$ and $R$ are the discrete total population and the total
number of species (richness) present.

Clone counts are commonly used in the theory of nucleation and
self-assembly~\cite{WATTIS1998,LAKATOS_JCP,LEIQI}, where all particles
are identical and $c_{k}$ represents the number of clusters of size
$k$. They are equivalent to `species abundance distributions' or
sometimes ambiguously described as `clone size distributions.' Clone
counts have \replaced{recently been}{been recently} used to quantify
populations in barcoding studies~\cite{goyal2015mechanisms} described
below.

Clone counts do not depend on the specific labeling of the different
types $i$ and do not contain any identity information. However, since
the common diversity indices are only a summary of the vector
$\{n_{i}\}$ and also do not retain species identity information,
$^{q}\!D$ can be written in terms of $c_{k}$ rather than $n_{i}$:
\begin{equation}
^{q}\!D = \left[\sum_{k=1}^{\infty}c_{k}\left({k \over N}\right)^{q}\right]^{1/(1-q)},
\end{equation}
which leads to corresponding expressions at specific values of $q$, \emph{e.g.}, 
$^{0}\!D = R$, 
\begin{eqnarray}
^{1}\!D = \exp\left[-\sum_{k=1}^{\infty} c_{k}\left({k\over
      N}\right)\ln \left({k \over N}\right)\right]\,\,\, \mbox{and} \nonumber \\
 1/{^{2}\!D} =  \sum_{k=1}^{\infty} c_{k}\left({k \over N}\right)^{2}.
\end{eqnarray}

While \deleted{the definitions of} $^{q}\!D$ \replaced{is}{are}
well-defined when \replaced{species are discretely delineated}{the
  discrete species are delineated}, for more granular or continuous
traits, the delineation of different species will affect the values of
$n_{i}$ and $c_{k}$. Fig.~\ref{CONFIGS} shows population counts
ordered by a continuous trait $x$. By defining the discrete species
$i$ according to different binning windows over $x$, we find different
sets of number and clone counts.  Thus, measures of diversity can be
highly dependent on the resolution and definition of traits and
species.
\section{Sampling}
\label{sec:SAMPLING}
In most applications, including all the ones we will discuss below,
the entire population is not accessible for identification and
measurement. In an \replaced{ecosystem}{ecology}, all animals of the population cannot be
tracked. In blood samples, only a small fraction of the cell types in
the whole organism \replaced{are}{is} drawn for identification/sequencing. Thus,
inferring the diversity in the entire system from the diversity in the
sample is a key problem encountered across many fields.

There are numerous ways to randomly sample a population. One approach
is to draw one individual, record its attributes, return it \replaced{to}{back into}
the system, and allow it to well-mix or equilibrate before again
randomly drawing the next individual. This process can be repeated $M$
times. To indicate this type of sampling, we use the subscript
$1\times M$ in the corresponding distributions and expectation
values. Similar sampling approaches are used in the
`mark-release-recapture' experiments to estimate population
size~\cite{mark_release_recapture_2017B}, survival, and dispersal of
mosquitos~\cite{mark_release_recapture_2017}. For a given
configuration $\{ n_{i}\}$ and total population size
$N$~\cite{mark_release_recapture_2013}, the probability that the
configuration $\{m_{i}\}$ is drawn after $M$ samples is simply
\begin{equation}
P_{\rm 1\times M}(\m\vert \n, M,N) = {M \choose m_{1}, m_{2}, \ldots,
  m_{R}}\prod_{j=1}^{R} f_{j}^{m_{j}},
\label{PS1}
\end{equation}
where $f_{j} \equiv n_{j}/N$ is the relative population of species
$i$, $N \equiv \sum_{i=1}^{R} n_{i}$ is the total 
population and $M \equiv \sum_{i=1}^{R}m_{i}$ is the total
number of samples.

We can now use $P_{\rm 1\times M}$ to compute the statistics of how
the system diversity is reflected in the diversity in the samples.
For example, the mean population in the sample in terms of $n_{i}$ is
$\mathds{E}_{1\times M}[m_{i}]\equiv \sum_{\m} \m P_{\rm 1\times
  M}(\m\vert \n, M,N)$.  The lowest moments of the populations
  in the sample are
\begin{eqnarray}
\mathds{E}_{1\times M}[m_{i}] =  M f_{i} = n_{i}{M \over N}, \\
\mathds{E}_{1\times M}[m_{i}m_{j}]  =  f_{i}f_{j}M(M-1) + f_{i} M\mathds{1}(i,j).
\nonumber 
\end{eqnarray} 
An alternative random sampling protocol is to draw a fraction $\sigma
\equiv M/N < 1$ of the entire population once. This type of sampling
arises in biopsies such as laboratory blood tests. To be able to
distinguish between this sampling protocol and the previous one, we
now use the notation $M\times 1$. In this case\added{,} the combinatorial
probability of a specific sample configuration, given $\n$, $N$, and
$M$ is
\begin{equation}
P_{\rm M\times 1}(\m\vert\n, M, N) = 
\prod_{j=1}^{R}\frac{{n_{j} \choose m_{j}}}{{N\choose M}}
\mathds{1}\!\left(\!M, 
\sum_{i=1}^{R} m_{i}\!\right),
\end{equation}
where the discrete indicator function enforces the constraint between
$m_{i}$ and the sampled population $M$. In this single-draw sampling
scenario, we use the Fourier decomposition $\mathds{1}(x,y) \equiv
\int_{0}^{2\pi}{\textrm{d} q \over 2\pi}e^{iq(x-y)}$ to find

\begin{eqnarray}
\mathds{E}_{M\times 1}[m_{i}] = n_{i}{M \over N} = n_{i} \sigma,
\\ \mathds{E}_{M\times 1}[m_{i}m_{j}] = n_{i}n_{j}{M \over N}{M-1\over
  N-1} \nonumber 
\\ \hspace{2.8cm} + \mathds{1}(i,j)n_{i}{M \over N}\left({N-M \over N-1}\right).
\end{eqnarray}
%
%
Results using $P_{\rm 1\times M}$ and $P_{\rm M\times 1}$ rely on
perfectly random sampling, where certain clones/species are not more
likely sampled or captured than others. The moments
$\mathds{E}[m_{i}m_{j}]$ can be directly used to evaluate the expected
Simpson's diversities, $S_{\rm r}$ (with replacement) and $S$ (without
replacement) defined by Eqs.~(\ref{SIMPSONINDEX}) and
(\ref{SIMPSONINDEX2}), in the corresponding sample. In the case of
$1\times$M sampling, we find
\begin{eqnarray}
\mathds{E}_{\rm 1\times M}[S_{\rm r}] & = 
\mathds{E}_{\rm 1\times M}\left[\sum_{i}\left({m_{i}\over M}\right)^{2}\right] \nonumber \\
\: & = {M(M-1) \over M^{2}}\sum_{i} f_{i}^{2} + {1 \over M}\sum_{i}f_{i} \nonumber \\
\: & = S_{\rm r}\left(1-{1\over M}\right)+{1\over M},
\end{eqnarray}
and 

\begin{eqnarray}
\mathds{E}_{\rm 1\times M}[S] & = \mathds{E}_{\rm 1\times M}\left[\sum_{i}{m_{i}\over M}
{m_{i}-1\over M-1}\right] \nonumber \\
\: & = \sum_{i}{\mathds{E}_{\rm 1\times M}[m_{i}^2]\over M(M-1)} - 
\sum_{i}{\mathds{E}_{\rm 1\times M}[m_{i}]\over M(M-1)}\nonumber \\
\: & = \sum_{i}f_{i}^{2} \equiv S
\end{eqnarray}
while for M$\times 1$ sampling, we find

\begin{eqnarray}
\mathds{E}_{\rm M\times 1}[S_{\rm r}] & = 
\mathds{E}_{\rm M\times 1}\left[\sum_{i}\left({m_{i}\over M}\right)^{2}\right] \nonumber \\
\: & = S_{\rm r} {M-1\over M-\sigma} + {1-\sigma \over M-\sigma}
\end{eqnarray}
and 

\begin{eqnarray}
\mathds{E}_{\rm M\times 1}[S] & = \mathds{E}_{\rm M\times 1}\left[\sum_{i}{m_{i}\over M}
{m_{i}-1\over M-1}\right] \nonumber \\
\: & = \left[\sum_{i}{\mathds{E}_{\rm M\times 1}[m_{i}^{2}]\over M(M-1)} -
\sum_{i}{\mathds{E}_{\rm M\times 1}[m_{i}]\over M(M-1)}\right] \nonumber \\
\: & = S.
\end{eqnarray}
Note that for both types of random sampling, we find that the expected
Simpson's diversity (without replacement) in the samples are equal to
the Simpson's diversity in the full system. In general, the
expectations do not commute and $\mathds{E}[S] \neq
S(\mathds{E}[m_{i}])$.

Effects of sampling on clone counts $c_{k}$ can be similarly
calculated by averaging the definition for the sampled clone count

\begin{equation}
b_{k}\coloneqq \sum_{i=1}\mathds{1}(m_{i},k) \in \mathbb{Z}^{+}
\label{DKEQN}
\end{equation}
over the sampling probabilities $P_{\rm M\times 1}(\m\vert\n, M, N)$
or $P_{\rm 1\times M}(\m\vert\n, M, N)$. For clone counts, the
calculations of moments of sampled quantities $b_{k}$ are more
involved\deleted{,} and explicitly noncommutative $\mathds{E}[b_{k}]
\neq \sum_{i}\mathds{1}(\mathds{E}[m_{i}],k)$. One advantage of
working in the $b_{k}$ representation is that diversity
indices\added{,} such as the expected sampled richness \added{$R^{\rm
    s}$,} are difficult to extract from $\mathds{E}[m_{i}]$ but
\replaced{are}{is} simply \added{found via} $\mathds{E}[R^{\rm s}]=
\sum_{k} \mathds{E}[b_{k}]$.  Some related results are given in
\cite{gotelli2013measuring,chao2014rarefaction}.


The above results provide expected diversities \textit{in the sample}
assuming full knowledge of $\{n_i\}$ in the system. They represent
solutions to the forward problem, the so-called \deleted{mean} `rarefaction'
in ecology. However, the problem of interest is usually the inverse
problem, or \emph{extrapolation} in ecology.  In the simplest case, we
wish to infer the expected diversity (or $\{n_{i}\}$ and $c_{k}$) in
the system from a given configuration $\{m_{i}\}$ or clone count
$b_{k}$. Extrapolation is a much harder problem and is the subject of
many research papers
\cite{FISHER1943,BUNGE_REVIEW1993,HSIEH2016,COX_HILL_2017,BUDKA2019}.

One may wish to use the observed sample diversity $^{q}\!D(M)$ to
approximate the population diversity $^{q}\!D(N)$.  For any $q$, the
underestimation of $^{q}\!D(N)$ using $^{q}\!D(M)$ decreases as the
sample size $M$ increases. The deviation of $^{q}\!D(M)$ from
$^{q}\!D(N)$ is smaller for larger $q$, as higher-order Hill numbers
are more heavily weighted by large species, which are less sensitive
to subsampling.

Chao and others have shown that for $q \geq 1$ and in the $N \to
\infty$ limit\added{,} nearly unbiased approximations can be
obtained\added{,} and when $q \geq 2$, these unbiased estimates are
very insensitive to sample size $M$
\cite{gotelli2013measuring,chao2014rarefaction}.
Using clone counts in a sample of population $M$,
Chao {\it et al.} \cite{chao2013entropy} obtained for $q=1$
(in terms of Shannon's index):
\begin{eqnarray}
&&
\hat{{\rm Sh}} = \sum_{k=1}^{M-1} \frac{1}{k}\!\sum_{1\leq m_i \leq M-k}\!\frac{m_i}{M} 
\frac{ {{M-m_i}\choose{k}} }{ {{M-1}\choose{k}} } 
\nonumber \\ && ~~
\hspace{2mm} - \frac{d_1}{M(1-A)^{M-1}} \left\{\log A +
\sum_{r=1}^{M-1} \frac{1}{r} (1-A)^r \right\},  
\end{eqnarray} 
where $A = {2 d_2} / {[(M-1)d_1 + 2d_2]}$. 

For $q \geq 2$, Gotelli and Chao \cite{gotelli2013measuring} obtained
\begin{eqnarray}
^{q}\!\hat{D} = \left[\sum_{m_i \geq q} \frac{m_i^{(q)}}{M^{(q)}}\right]^{1/(1-q)}
\end{eqnarray}
where $x^{(j)} = x(x-1)...(x-j+1)$.  For example, $^{2}\!\hat{D} =
M(M-1) / \sum_{m_i \geq 2} m_i (m_i-1)$, the inverse of Simpson's
index without replacement (Eqs.~\ref{SIMPSONINDEX0} and
\ref{SIMPSONINDEX2}). 

The ill-conditioning of the inverse problems is particularly severe
for the richness $^{0}\!D$.  The general formula for an estimate of
the system richness is
\begin{eqnarray}
^{0}\!\hat{D} = R(M) + \hat{d}_0,
\end{eqnarray}
and reduces to the unseen species problem for determining $d_{0}$
\cite{EFRON1976,UNSEEN_SPECIES2016}.  Since the sample size $M$ and
the richness $R$ in the system are uncorrelated, \deleted{rigorously,} one must
use information contained in the species fractions $f_{i}$ or the
clone counts $c_{k}$ in the full system \cite{WILLIS2015,WILLIS2016}.
However, a popular estimate for the system richness $R(N)$ is the
`Chao1' estimator \cite{chao1984nonparametric,gotelli2013measuring}

\begin{eqnarray}
{\rm Chao1}:\, \hat{R}(N) = R(M) + \frac{d_1^2}{2d_2},
\end{eqnarray} 
which is actually a lower bound and gives reliable estimates for
systems of size only up to approximately double or triple the sample
size $M$.  The uncertainty of the Chao1 estimator has also been
derived via a variance that is also a function of $d_{1}$ and $d_{2}$
\cite{chao1987estimating}.  The `Chao2' estimator gives the system
richness as a function of measured incidence
\cite{gotelli2013measuring}

\begin{eqnarray}
{\rm Chao2}:\, \hat{R}(N) = R(M) + \frac{q_1^2}{2q_2},
\end{eqnarray} 
where $q_{1}, q_{2}$ are the number of species found in 1 or 2 samples
out of many (as in the $1\times M$ sampling method).  Shen \textit{et
  al.}  \cite{shen2003predicting} derived another estimate

\begin{eqnarray}
\hat{R}(N) = R(M) + d_0
\left[1 - \left(1 - \frac{d_1}{M d_0 + d_1} \right)^{\!N-M} \right], 
\end{eqnarray}
which is only reliable if the sample size $M$ is more than half of the
system size $N$. Many of these estimators have been coded into
analysis software such as R and iNEXT
\cite{hsieh2016inext}.

Regardless of the estimator, the major limitation is an insufficient
sample size $M \ll N$.  Models predicting species abundances as a
function of system size can help bridge this gap.  For
example\added{,} as log-normal relationship for the clone count $c_{k}$
\cite{CURTIS2002} has been used to find agreeable results
\cite{LENNON2016,LENNON2016B}.  In general, models can be extremely
useful \replaced{for quantifying}{analyzing} the effects of sampling,
particularly when a Bayesian prior is desired.


%
%
%
%

We have outlined the basic mathematical frameworks for quantifying
diversity that have utility across applications in different
disciplines. The above summary of sampling assumes a \emph{well-mixed}
population, precluding any spatial dependence of the distribution of
individual species. Spatially dependent sampling has been proposed for
the origin of relationships between the number of species detected and
the total area occupied by the population (see below).  

\section{Fields in which diversity play a key role}
\label{sec:FIELDS}

Below, we summarize a few modern applications in which diversity is
important. By no means exhaustive, the following are simply examples
of specific systems in \deleted{modern} biology that reflect the authors'
intellectual biases.

\subsection{Ecology, paradox of the plankton}
The classic problem in the context of biological diversity is dubbed
\emph{the paradox of the plankton} and was originally discussed in a
paper of the same title~\cite{hutchinson1961paradox}. It describes
diverse populations of plankton in environments \replaced{with
  limited}{of limited number of} resources or nutrients. Sampled
populations of plankton exhibit a large number of species even in low
nutrient conditions during which one expects strong competition for
resources.  This observation runs counter to the \emph{competitive
  exclusion principle} arising in many
settings~\cite{hardin1960competitive}.

Perhaps the most common application of diversity arises in biological
population studies, specifically in
ecology~\cite{fisher1943relation,magurran1988ecological,
  benton1995diversification,courtillot1996effects,alroy2004sepkoski,stollmeier2014possible}. Possible
areas of application include the monitoring of ecosystems and the
development of efficient species conservation
strategies~\cite{heywood1995global,courtillot1996effects,john1998rates,sala2000global,margules2000systematic,alroy2004sepkoski,sarkar2006ecological}.
Multiple overlapping and nebulous definitions of ecological diversity
have been advanced
\cite{simpson1949measurement,hurlbert1971nonconcept,sepkoski1988alpha,purvis2000getting,WHITTAKER2001,ricotta2005through,sarkar2006ecological}.
Early work by Fisher~\cite{fisher1943relation} introduced a
logarithmic series model to mathematically describe empirical species
diversity data. Here, the diversity index referred to a free parameter
in the corresponding model.  In a later study, MacArthur defined
species diversity based on the size of the sampled
area~\cite{MACARTHUR1965}.  In the ecological setting, multiple layers
of subpopulations are an important feature of populations. These
subpopulations may be delineated by another property of the individual
species, such as size, weight, behavioral attributes,
etc. Subpopulations can also be distinguished through their spatial
distribution or occupation of different
habitats. Whittaker~\cite{WHITTAKER1960,WHITTAKER1977} qualitatively
defined four types of diversity (point, alpha, beta, and gamma)
conditioned on habitat or spatial distribution of the
subpopulations~\cite{WHITTAKER1977}. Fundamentally, these differences
arise from different methods of sampling, leading to different Hill
numbers $^{q}\!D$. We summarize a few often-used descriptions below:

\begin{itemize}

\item `Point diversity' refers to samples taken at a single point or 
`microhabitat.' This quantity is usually operationally measured 
by trapping organisms at one or more specific points.

\item `Alpha diversity' is defined as the diversity within an
  individual location or specific area. In general, one can define a
  Hill number derived from measurements at a specific location as
  $^{q}\!D_{\alpha}$, while the index $\alpha \equiv
  {^{0}\!D_{\alpha}}$ is the richness encountered within a defined area
  or specific location. 
%
%
A few subtle variations in the definition of the index $\alpha$ exist,
mostly related to the sampling
process~\cite{TUOMISTO2010B,jost2007partitioning}. For example, in
relation to beta diversity (discussed below), alpha diversity is the
mean of the specific-location diversities across all locations within
a larger landscape.


\item `Gamma diversity' is the diversity index $^{q}\!D_{\gamma}$
  determined from the entire dataset, the total landscape, or
  \added{the} entire ecosystem. The index $\gamma \equiv
        {^{q}\!D_{\gamma}}$ usually denotes the total number of
        different species or clones at the largest scale. Note that
        the mean or sum of the alpha diversities is in most cases not
        equal to the gamma diversity. The nonlinearity of the Hill
        numbers as well as the intersection or exclusion of species
        amongst the different sites suggests a need for indices that
        connect alpha and gamma diversities.


\item `Beta diversity' was devised to describe the difference in
  diversity between two habitats or between two different levels of
  ecosystems. While the different levels of diversity are designed to
  the spatial aspects of diversity, different habitats overlap,
  leading to some amount of arbitrariness in determining
  \deleted{sampling of} the $\beta$-diversity. Moreover, beta
  diversity was initially described in different ways
  \cite{WHITTAKER1960,WHITTAKER1977,TUOMISTO2010B}, leading to
  confusion about its mathematical definition and
  use~\cite{jost2006entropy,jost2007partitioning,TUOMISTO2010B}.  One
  possible definition is Whittaker's \cite{WHITTAKER1960}
  multiplicative law ${^{q}\!D_{\gamma}}\equiv{^{q}\!D_{\alpha}}
  {^{q}\!D_{\beta}}$ where here, $\alpha$ is defined as the
  \textit{mean} of the diversities across all
  micro-habitats. Whittaker's definition describes beta diversity
  ${^{q}\!D_{\beta}}={^{q}\!D_{\gamma}} /{^{q}\!D_{\alpha}}$ as a
  measure to quantify the diversity in the total population relative
  to the mean diversity across all
  micro-habitats~\cite{TUOMISTO2010B}. In the limit of $q\rightarrow
  1^-$, we obtain the Shannon diversity relationship ${\rm
    Sh}_{\gamma}={\rm Sh}_{\alpha}+{\rm Sh}_{\beta}$ according to
  Eq.~(\ref{SHANNONDIV}). Another definition of $\beta$ is given by
  Lande's \cite{lande1996statistics} additive law $\gamma \equiv
  \alpha + \beta$ according to which diversity indices are measured in
  the same units. One concept associated with $\beta$ in terms of the
  additive partitioning is `species turnover,' quantifying the
  difference in richness between the entire and the local population.
  As an example, consider two distinguishable or spatially separate
  habitats \texttt{A} and \texttt{B}. If \texttt{A} contains species
  $\{a,b,c,d,e\}$ and \texttt{B} contains $\{b,c,f,g\}$, we find
  $\beta_{ \texttt{A},\texttt{B}} = 5$ associated with the set
  $\{a,d,e,f,g\}$.  The laws of Whittaker and Lande sparked debates
  about how to properly define beta diversity, and led to the
  distinction between multiplicative and additive diversity
  measures~\cite{jost2006entropy,jost2007partitioning,TUOMISTO2010B}.

%
%

\item `Delta, Epsilon, Omega diversity' are other hierarchical
  definitions of diversities proposed by Whittaker
  \cite{WHITTAKER1977}. Delta diversity is analogous to beta diversity
  but defined at the larger among-landscape scale, while epsilon
  diversity corresponds to gamma diversity, but at the \emph{regional}
  scale that contains many \emph{landscapes}.  Omega diversity is
  measured at the biosphere scale, and thus characterizes the
  diversity of all ecosystems~\cite{contoli2015contributions}.

\item `Zeta diversity' was introduced by Hui and McGeoch
  \cite{ZETA2014}, and is defined by a set of $\zeta$ indices that
  mathematically describe the species numbers between different
  partitions of a certain habitat. Specifically, $\zeta_{i}$ is the
  mean number of species shared by $i$ partitions. In particular,
  $\zeta_{1}$ is the mean richness across all sites. For example,
  between two samples $\texttt{A}$ and $\texttt{B}$ or sets of data,
  the average number of species is $\zeta_{1}\coloneqq
  (R_{\texttt{A}}+R_{\texttt{B}})/2$, while the intersection is
  $\zeta_{2} \coloneqq \texttt{A}\cap\texttt{B}$. Generalizations to
  multiple samples can be defined using a series of zeta diversity
  indices $\zeta_{i}$.

\item Many other indices have been defined for different applications. The
  Jaccard
  index~\cite{jaccard1900contribution,WHITTAKER1960,TUOMISTO2010B,ZETA2014}
  is defined as $J(\texttt{A},\texttt{B})=\vert \texttt{A} \cap
  \texttt{B}\vert/\vert \texttt{A} \cup \texttt{B}\vert$, and is a
  general measure for quantifying the similarity in richness between
  two sets of populations $\texttt{A}$ and $\texttt{B}$.  Margalef's
  index~\cite{margalef1957information} and Menhinick's
  index~\cite{menhinick1964comparison} are relative richness measures
  given by $R/\ln N$ and $R/\sqrt{N}$, respectively. Other indices
  include the Bray-Curtis dissimilarity~\cite{bray1957ordination}, the
  Berger-Parker diversity index~\cite{berger1970diversity} as defined
  in Eq.~(\ref{BERGERPARKER}), Fager's
  index~\cite{fager1957determination}, Keefe and Bergersen's
  index~\cite{keefe1977simple}, McIntosh's
  index~\cite{mcintosh1967index}, and Patil and Taillie's
  index~\cite{patil1982diversity}.
\end{itemize}

A myriad of different definitions of diversity indices arise from
specific cases of the Hill numbers and consideration of different
spatial scales of ecosystems. There is potential to further unify
these definitions in a more systematic way using mathematical norms
and more general mathematical structures of spatial dispersal of
particles.


%
%
%
\subsection{Area-Species Law and Island Biodiversity}
A particularly consistent, albeit qualitative feature observed in
ecology is the species-area relationship (SAR) which relates the
measured number of species (richness) \replaced{to}{with} the relevant area. These
areas can represent distinct habitats, such as mountain tops, or islands. For
the latter, much work has been done in the subfield of island biodiversity.

\begin{figure}
\centering
\includegraphics[width=2.6in]{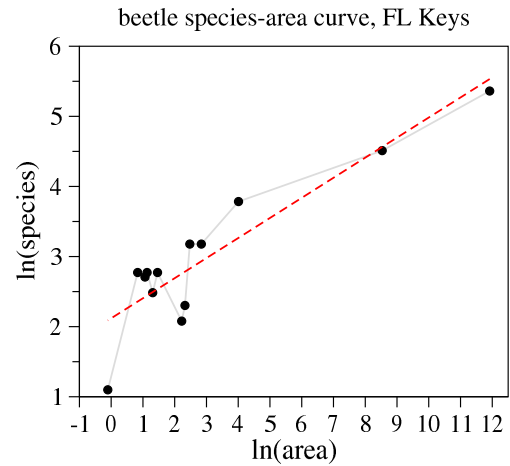}
\caption{Plot of $\ln R$ versus $\ln A$ with area $A$ measured in
  terms of km$^{2}$. Species counts of long-horned beetles in the
  Florida Keys are plotted against the island size \cite{KEYS1996}. The
  linear regression line yields a slope of $z=0.29$. Usually, fits
  of the species-area exponent $z$ yield a small number.}
\label{AREA_SPECIES}
\end{figure}

The SAR is usually expressed as a power-law relationship between the
number of species (or richness) $R$ and the habitat/island area:
\begin{equation}
R = c A^{z},
\label{SAR_EQN}
\end{equation}
where $c$ is a constant prefactor and $z$ is an exponent.  On a
log-log plot, $\log R = \log c + z\log A$ defines a line with slope
$z$. An example of the area-species law for species counts of
long-horned beetles in the Florida Keys is shown in
Fig.~\ref{AREA_SPECIES}, yielding a slope $z=0.29$. An alternative
species-area relationship is $e^{R} = cA^{z}$ \cite{GLEASON}, which is
a straight line on a semi-log plot.

The classic book by MacArthur and Wilson \cite{ISLANDBIO1967} and many
subsequent analyses have promoted and extensively analyzed the SAR
idea. In MacArthur and Wilson's neutral equilibrium theory, immigration to and
death on an island are monotonically decreasing and increasing
functions of the number of species already on the island,
respectively.
Usually, measured values of the exponent fall in the range $z \sim
0.1-0.4$.  Field work has also found relationships between the
parameters $c$ and $z$ and system-specific attributes such as the
island distance to the mainland, habitat type, etc
\cite{ISLANDBIO1967,ISLANDBIO2003}. Nonetheless, reasonable
predictions based on Eq.~(\ref{SAR_EQN}) are ubiquitous across many
ecological examples.

Mechanistic origins of the robustness of the SAR have been proposed
\cite{MCCOY1979,HE_LEGENDRE2002,GOLDENFELD2006}. Different models for
species populations $n_{i}$ or clone counts $c_{k}$ were surveyed\added{,} and
the corresponding species-area laws were derived by He and Legendre
\cite{HE_LEGENDRE2002}. Spatial clustering of species and the
averaging of random measurements \replaced{were}{was} shown to robustly generate a
power-law species-area curve \cite{HE_LEGENDRE2002,GOLDENFELD2006},
highlighting the fundamental importance of sampling.

\begin{figure*}[htp!]
\centering
\includegraphics[width=6.5in]{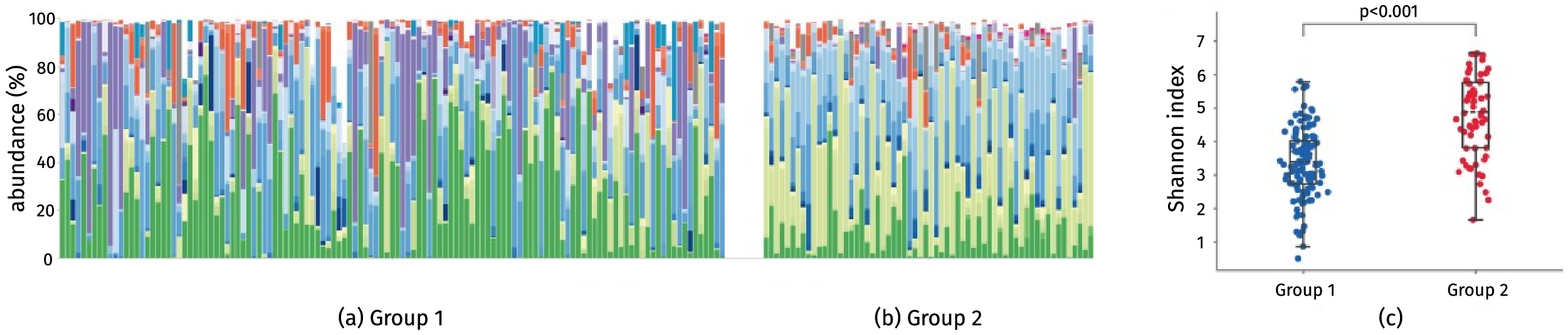}
\caption{Frequencies of approximately 200 species of bacteria
  distributed across about a dozen phyla. (a) Group 1 depicts the
  relative \added{species} abundance distribution \replaced{in samples
    from patients with Crohn's disease}{for healthy individuals} while
  (b) Group 2 shows the \added{species} pattern \replaced{in normal
    patients}{for irritable bowel syndrome (IBD) patients}. The
  differences in abundance patterns are apparent and have been
  quantified using the Shannon index for each individual plotted in
  (c). From Park \textit{et al}.~\cite{HYLEE2019}. }
\label{GUT_SPECIES}
\end{figure*}
\subsection{Gut Microbiome}

Another ecological system that has recently received much attention is
the human microbiome, especially in the gut. The gut bacterial
ecosystem is important for health and can impact cardiovascular
disease, diabetes, neuropsychiatric diseases, inflammatory bowel
disease (IBD), \added{and} digestive and metabolic function to the point that
fecal transplantation (bacteriotherapy) has become an effective
treatment for recurrent \emph{C.~difficile colitis}
infections~\cite{FECALTRANSPLANT2019}. This type of infection often
occurs after antibiotics disrupt the gut microbiome. Transplants have
also shown to be effective in treating slow-transit
constipation~\cite{CONSTIPATION2017}.

Recent efforts to collect and curate gut microbiome data have included
NIH's Human Microbiome Project (HMP)~\cite{NIH2014,NIHURL} and the
European Metagenomics of the Human Intestinal Tract
(MetaHIT)~\cite{NATURE_GUT_EU,METAHIT,METAHITURL}, as well as the
integration of the data in~\cite{INTEGRATED2014}. Each
dataset contains sequence data from samples from different body
regions of hundreds of individuals, both healthy and diseased.

Bacterial species are usually determined by sequencing of the 16S
ribosomal RNA (rRNA), a component of prokaryotic ribosomes that
contain hypervariable regions that are species-specific.  However,
closely related taxa can have very similar sequences, making
separation imperfect~\cite{BALDRIAN2013}. Nonetheless, with numerous
public databases~\cite{METABOLOMICS,CME_MSU,EZBIOCLOUD,HYLEE2019},
estimates of species abundances in samples are readily available. In
the gut, there are usually on the order of $10^{3}$ bacterial species,
with Bacteroidetes and Firmicutes being the dominant
phyla~\cite{GUT_REVIEW2015,THURSBYREVIEW}. Indeed, lower gut diversity
is seen to be associated with conditions such as Crohn's
disease~\cite{GUT_REVIEW2015}. For example, the frequency distribution
of bacterial species \replaced{from}{in} healthy and \replaced{Crohn's
  disease}{irritable bowel syndrome} patients are shown in
Fig.~\ref{GUT_SPECIES}.  The quantification of diversity of human
microbiome is an essential step in ongoing research\added{,} and
\deleted{the} diversity indices have been applied to microbiome data,
including $\alpha$-diversity and $\beta$-diversity across the
microbiome from different anatomical regions and different
patients. As with island biodiversity, the gut microbiome can be
modeled as \added{a} birth-death-immigration (BDI) process.

\subsection{Barcoding Experiments}
\begin{figure*}
\centering
\includegraphics[width=5.4in]{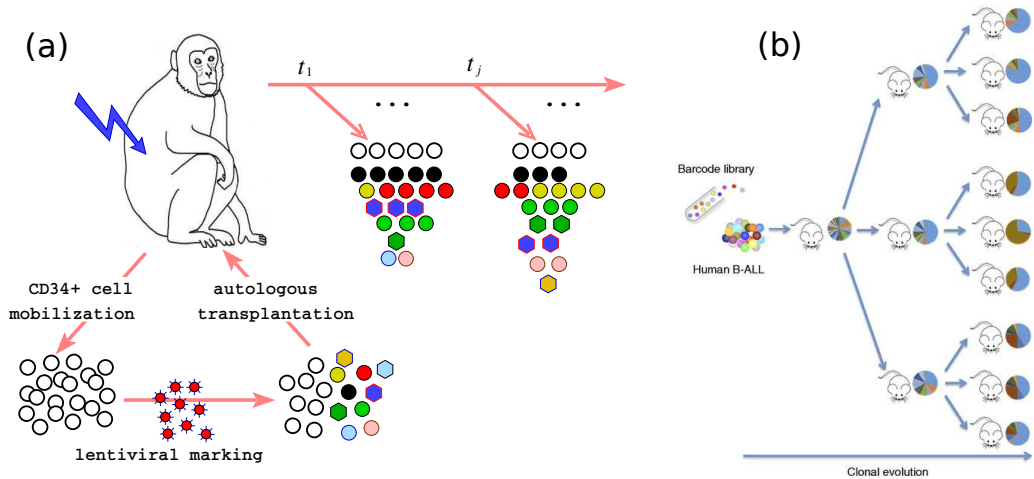}
\caption{(a) Protocol for Viral Integration \replaced{Site}{site}
  (VIS) \replaced{barcode}{barcoding} studies of hematopoiesis in
  rhesus macaque
  \cite{kim2014dynamics,dunbar2014,goyal2015mechanisms}. Here,
  `barcodes' are defined by the random integration sites of a
  lentiviral vector.  (b) Xenograft barcode experiments using mice
  \cite{Bystrykh2017} in which a library of barcodes was used to tag
  leukemia-propagating cells before direct transplantation into mice.}
\label{BARCODE_EXPT}
\end{figure*}
Besides taxonomy of gut bacteria, the accurate identification of
animal and plant species from samples is an essential task in
ecology. In the early 2000's\added{,} a DNA barcoding method was
developed to read relatively short DNA regions specific to certain
species~\cite{HEBERT2003,HEBERT2004}. These \textit{barcodes} are
usually found in mitochondrial DNA and often derived from a region in
the cytochrome oxidase gene~\cite{HEBERT2003}. By sequencing samples
and comparing \added{them} with a sequence database such as The
Barcode of Life Data System~\cite{HEBERT_BOLD,BARCODEOFLIFE}, one can
infer the number of species present within a sample.  Detecting
specific species within samples using DNA barcoding and DNA libraries
\replaced{arises}{have been used} in many applications including
identification of birds \cite{HEBERT2004} \replaced{and}{,
  identification of} flowering plants~\cite{KRESS2005},
\replaced{detection of}{detecting} contaminants~\cite{SGAMMA2018}, and
\added{the} tracking \added{of} plant composition in processed
foodstuffs~\cite{BRUNO2019}.

Recently, a number of barcoding or tagging protocols
\cite{INDEL_PNAS,LIMITATIONS_SCI_REP,TAMBE2019} have been developed to
genetically label a large population of cells to study how they
differentiate and proliferate, especially in the context of
hematopoiesis
\cite{sun2014clonal,kim2014dynamics,dunbar2014,perie2016retracing}
and cancer progression
\cite{BLUNDELL2014,BARCODE_CANCER2018,AKIMOV2019}.
%
%

A novel approach used to investigate hematopoiesis exploits \textit{in
  situ} barcodes~\cite{sun2014clonal}. Mice were engineered with an
enzyme (Sleeping Beauty Transposase) that randomly moves DNA sequences
(transposons) to different parts of the genome. The transposase is
designed to be controllable by doxycycline, an antibiotic that can be
used to switch on or off gene regulation.  When the transposase is
briefly activated, transposons within cell \added{genomes} are
randomly rearranged within a brief period \added{of time}.  Since the
genome length $\gg$ transposon length, the new locations of the
transposons will be distinct across the \emph{founder} cells. After
switching off the transposase, proliferation of founder cells
\deleted{will} \replaced{imparts}{impart, except for rare DNA
  replication events,} the same genomic sequence to
\replaced{their}{its} daughter cells.  These collections of cells
constitute a multiclonal population that proliferates and
differentiates.

Analysis of the clonal population within differentiated cell pools
show\added{s} that granulocytes derive from stem cells at particular
time points during the life of the mouse \cite{sun2014clonal}.
Comparing clonal abundance structure within different cell lineages
\replaced{shows}{showed} that clones originally predominant in the
lymphoid lineages eventually arise in myeloid cells, indicating that
multipotent progenitor cells continually produce cells of both
lineages.  \deleted{These conclusions arise after statistical analysis
  of the clone (defined by their transposon sites) abundance
  distribution within different groups of cells.}


In another recent series of studies on hematopoiesis, \added{outlined
  in Figure~\ref{BARCODE_EXPT}}, stem cells (HSCs) were extracted from
rhesus macaques and infected with a lentiviral vector. The lentivirus
integrates its genome randomly in the genome of the HSCs. Since the
lentivirus genome is much shorter than that of mammalian cells, nearly
every successful infection results in a new viral integration site
(VIS) or clone.  The infected stem cells are autologously transplanted
into the animal\added{,} and some of them resume differentiation into
progenitor cells that transiently proliferate and further
differentiate. Descendant cells carry the same genetic sequence,
including the lentivirus integration locations, or the viral
integration sites (VIS).  Another approach is to use libraries of
synthesized DNA/RNA as tags. Here, the different sequences, rather
than their integration sites, serve as the distinguishing
feature. This process avoids the need to determine VISs.

In all \added{of} the above approaches, each successive generation of cells will
acquire the same tag, VIS or specific DNA barcode sequence\deleted{,} as their
parent\added{,} and ultimately\added{, as} the founder HSC.  Compared to the Sleeping
Beauty Transposon protocol, the VIS or barcoding experiments require
an additional viral transfection step. Nonetheless, these VIS and
barcoding experiments are equally effective in dissecting the
differentiation process and quantifying lineage bias with age.  For
example, the variation (in time) of the abundances of a clone across
different lineages indicates the level of fate switching of a stem
cell~\cite{kim2014dynamics,koelle2017quantitative}.

These experiments also enabled observation of biological mechanisms on
a finer scale compared to traditional studies, allowing inference of
parameters that are difficult to measure directly such as the initial
HSC differentiation rate and the proliferative potential (number of
generations) accessible to progenitor
cells~\cite{goyal2015mechanisms,xu2018modeling}.

\begin{figure}[h!]
\centering
\includegraphics[width=2.7in]{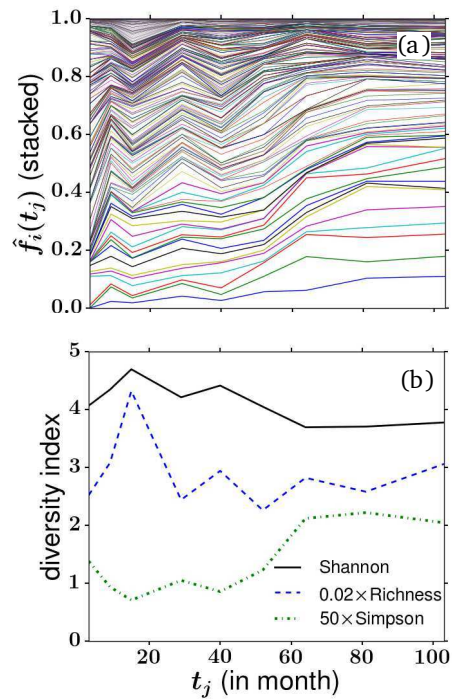}
\caption{(a) The fractional populations of the largest clones
  (barcodes) detected in granulocyte blood samples from rhesus
  macaque. Relative populations are described by the distances between
  neighboring curves. (b) Diversity indices derived from the data in
  (a). The Simpson's index and Shannon diversity are rescaled to fit
  on the same plot.}
\label{CK_TCELL}
\end{figure}

After sampling, PCR amplification, and sequencing (each process
\replaced{carrying}{exhibiting their} specific errors), the relative
species populations and clone counts within defined cell types can be
quantified. Fig.~\ref{CK_TCELL}(a) shows frequencies of barcode $i$ as
a function of sampling times $t_{j}$ in rhesus macaque. The fraction
of each clone is depicted by the vertical distance between two
neighboring curves.
%
%
Here, it is important to note that the `diversity' is a measure of
the distribution of clone ID (barcodes) instead of lineages (cell
types). In Fig.~\ref{CK_TCELL}(b), we plot three different and
rescaled diversity indices associated with the data in (a). The
sampled richness is initially low at month 3 when barcoded clones have
not fully differentiated and emerged in the peripheral blood. The
sampled richness then peaks at month 9 before stabilizing after month
29. Simpson's diversity seems to continue to increase after month 29\added{,}
which may indicate more unevenness and coarsening (fewer clones
dominating the total population). Shannon's index is shown to decrease
slightly, suggesting a decrease in the effective number of barcodes.

Sun \textit{et al.} \cite{sun2014clonal} and Kim \textit{et al.}
\cite{kim2014dynamics} also used simple clustering algorithms that
identified similar clones according to their activity patterns across
time.  They identified distinct groups of clones that are featured by
different time points of contribution to hematopoiesis.  Koelle
\textit{et al.}~\cite{koelle2017quantitative} calculated Shannon
diversity to ensure comparability \added{across time} between animals
\added{and} different cell types\deleted{, and across time}.

\begin{figure}
\centering
\includegraphics[width=2.5in]{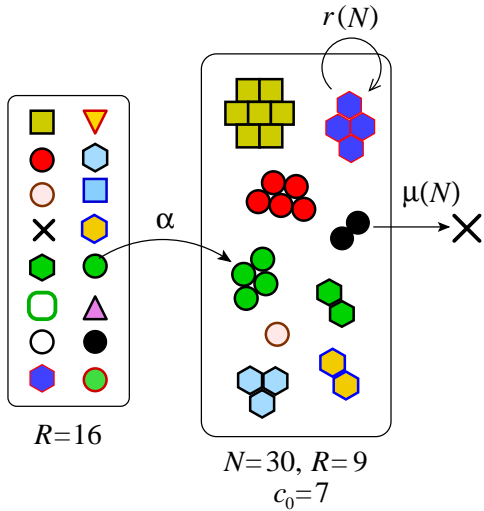}
\caption{A simple multispecies birth-death-immigration (BDI) process
  \cite{goyal2015mechanisms,xu2018modeling,dessalles2018exact,xu2018mathematical}. A
  constant source (\textit{i.e.}, stem cells with slow dynamics)
  \replaced{generated by}{of} 16 cells, each of a different clone,
  undergo asymmetric differentiation with rate $\alpha$ to produce
  differentiated cells that can undergo birth or death with rates
  $r(N)$ and $\mu(N)$ that may depend on the total population in the
  differentiated pool. In this example, the differentiated population
  contains $N=30$ cells, $R=9$ different clones (barcodes), thus
  leaving $c_{0}=7$ unseen species.}
\label{BDI}
\end{figure}
The employment of neutral barcodes to study blood cell populations is
statistically insensitive to spatial partitioning (different tissues
in the organism).  Nonetheless, small sampling \added{($M\ll N$)}
makes inference difficult.  Thus, mechanistic simplifications and
mathematical models have been used to quantify clonal
evolution. Assuming a multispecies birth-death-immigration process
(Fig.~\ref{BDI}) Dessalles \textit{et al.}~\cite{dessalles2018exact}
found explicit steady-state distribution functions for $n_{i}$ (log
series) and $c_{k}$ (Poisson) for constant $r$ and $\mu$, as well as
formulae for the expected Shannon's and Simpson's diversities. Goyal
\textit{et al.}~\cite{goyal2015mechanisms} derived a master equation
for the evolution of $\mathds{E}[c_k]$ and then extended the solution
to expected clone counts in the progenitor cell and sampled
\added{mature} cell pools.  By comparing results to the expected clone
count in the sample at steady-state, they were able to infer kinetic
parameters of the differentiation process. Biasco \textit{et
  al.}~\cite{biasco2016vivo} proposed two candidate stochastic models
for $n_i$ and used Bayesian Information Criterion (BIC) to assess the
likelihood of each.

\subsection{Cells of the Adaptive Immune System}
Another intra-organism system for which diversity is often quantified
is the adaptive immune system in vertebrates. The simplest immune
subsystem consists of lymphoid cells (\textit{e.g.}, B and T cells)
and tissues. B and T cells originate from common lymphoid progenitors
(CLPs) that differentiate from HSCs in the bone marrow. B cells
develop from CLPs in multiple stages in the bone marrow and spleen\added{,}
while T cells are formed from CLPs in the thymus.
%
%
During T cell development in the thymus, T cell receptors (TCRs) are
generated by random recombination of the associated receptor
gene. TCRs are heterodimeric proteins that usually consist of an alpha
chain and a beta chain.  After a specific genetic
sequence--corresponding to a specific amino acid sequence--is
\replaced{selected}{chosen and then selected for}, the \textit{naive}
T cell is exported from the thymus into peripheral tissue (such as
circulating blood and lymph nodes) where they can further proliferate
or interact with antigens presented on the surface of
antigen-presenting cells (APCs)\deleted{and become activated}.  Naive
T cells (those that have not previously strongly interacted with an
antigen) can be activated through association of the surface T cell
receptors (TCRs) with antigens presented by major histocompatibility
complex (MHC) molecules on the surface of APCs. Similarly, naive B
cells are generated in the bone marrow. The B cell receptors (BCRs)
are comprised of heavy and light chains and an antigen-binding region,
which is generated by the same recombination processes as TCRs.  B
cells are subsequently activated within tissues by binding to an
antigen via their B-cell receptors (BCRs).

The mechanism responsible for creating very diverse repertoires of
both BCRs and TCRs is V(D)J recombination~\cite{alt1992vdj}. In
developing B cells, this mechanism involves the random recombination
of diversity (D) and joining (J) gene segments of the heavy chain (DJ
recombination). In the following step, a variable (V) gene segment
joins the previously formed DJ complex to create a VDJ segment. In
light chains, D segments are missing and therefore only VJ segments
are generated. During T cell development and TCR generation, gene
segments of the alpha chain and beta chain, the VJ and VDJ segments,
respectively, also undergo random recombination. In the case of the
beta chain, one of two different D regions of thymocytes recombine
with one of six different joining J regions first, followed by
rearrangement of the variable V region connecting it to the
now-combined DJ segment. Due to the missing D segments in alpha
chains, only VJ recombination is taking place. The recombination and
joining processes in B cells and T cells involve many different
genetic deletions and insertions that result in many different BCR and
TCR protein sequences and a very large theoretical total number of
possible clones with $R \gtrsim
10^{14}-10^{15}$~\cite{LYTHE2016,Zarnitsyna2013}.

In the end, each T or B cell expresses only one TCR or BCR type (an
`immunotype' or `clonotype'). TCR sequences are preserved during
proliferation, while BCR sequences can further evolve \cite{PYBUS}.
Since the space of antigens (the different amino acid sequences, or
epitopes, presented by MHCs) is large, a large number of different TCR
and BCR sequences should be present in \replaced{an}{the} organism in order to mount
an effective response to a wide range of infections. However, before T
cell export from the thymus, a complex selection process
occurs~\cite{YATES2014}. Positive selection eliminates T cells that
interact too weakly with MHC molecules.  Subsequently, negative
selection eliminates those T cells and TCRs that bind too strongly to
epitopes. Cells that escape negative selection may lead to autoimmune
disease as they react to self-proteins. Thus, the total number of
different distinct immunoclones realized in an organism (the richness)
defines its T cell repertoire and is estimated to range from $10^{6} -
10^{8}$~\cite{MICEC}, with the lower range describing mice and the
higher range an estimate for humans. B cell richness in man is
estimated to be
$10^8-10^9$~\cite{dewitt2016public,rosenfeld2018computational}. These
values are much lower than the theoretical repertoire size $R \gtrsim
10^{14}-10^{15}$.  TCR and BCR diversity is an important factor in
health. For example, TCR diversity \deleted{it} has been shown to
influence the tumor microenvironment and \added{lymphoma patient}
survival\deleted{in lymphoma}~\cite{LYMPHOMA2017}.

\begin{figure}[h!]
\centering
\includegraphics[width=3.3in]{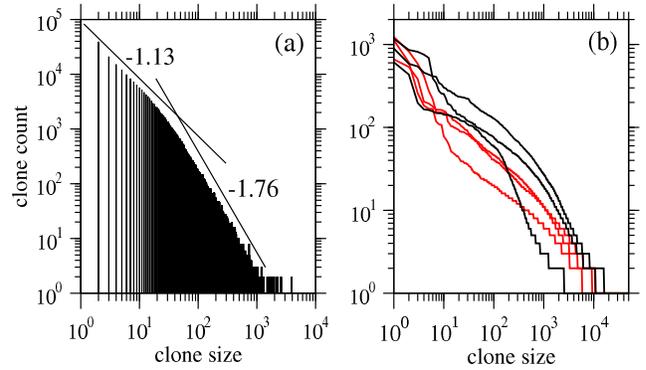}
\caption{Examples of recently published clone count data.  (a) Clone
  counts derived from a small sample ($10^{5}$ sequences) of T
  cells~\cite{Zarnitsyna2013}. Note the broad distribution described
  by a biphasic power-law curve. Ignoring the largest clones,
  power-law fits for each regime yield slopes of -1.13 and -1.76.
  However, one should be cautious describing sampled TCR (and
    BCR) clone counts using power laws as they hold typically for far
    less than two decades. (b) Human TCR clone counts for three
  HIV-infected (red) and three uninfected (black) individuals show
  qualitative differences between the distributions
  (unpublished). Other data from mice and humans, under different
  conditions and in different cell types, have been recently published
  \cite{SETHNA2017,Oakes2017}.}
\label{CLONEDATA}
\end{figure}


Although specific TCR sequences $i$ can be determined, and their
populations $n_{i}$ measured and estimated, the TCR identities vary
significantly across individuals (private sequences) so clone counts
are usually studied. Fig.~\ref{CLONEDATA}(a) shows T cell clone counts
$b_{k}$ sampled from mice~\cite{Zarnitsyna2013} that exhibit a
biphasic power-law behavior. Fig.~\ref{CLONEDATA}(b) shows preliminary
clone counts for six individuals, three
\replaced{uninfected}{HIV-negative} patients and three HIV-infected
patients \cite{OTTO}.

Quantifying T cell diversity is confounded by a number of technical
limitations. Usually, the complete T cell repertoire in an animal
cannot be directly measured. Rather, as in most other applications,
small samples of the entire population are usually drawn. \replaced{When}{In} sampling
from animals, the fraction of cells drawn and sequenced is perhaps
only $M/N \sim 10^{-5} - 10^{-2}$. Thus, clones that have small
populations may be missed in the sample. Besides sampling, sequencing
requires PCR amplification of the sample, leading to PCR bias,
especially in the larger-sized clones \cite{Oakes2017}.  Finally, as
in many other applications, there are multiple subclasses of the T
cell population. Naive T cells that are activated by antigens develop
into memory T cells that carry the same TCR and \deleted{that} can
further proliferate.  Thus, it is difficult to separate the clone
counts of different subpopulations such as naive or memory T cells
\cite{Oakes2017}.

Many mathematical models for the development and maintenance of the
immune systems have been developed
\cite{YATES2014,LYTHE2016,LYTHE2018,
  dessalles2018exact,xu2018modeling,xu2018immigration}.  For the
multiclonal naive T cell population, rudimentary insights can also be
gleaned from a birth-death-immigration process, much as in the
modeling of hematopoiesis. Here, the thymus mediates the immigration
of a large number of clones, which undergo homeostatic proliferation
and death in the periphery. Immigration rates can be different for
different clones, depending on the likelihood of specific
recombination patterns which may be inferred from probabilistic models
of VDJ recombination~\cite{Marcou2018,OLGA2019}.

Proliferation in the periphery depends on interactions between
self-peptides with T cell receptors and is thus
clone-dependent. Recently, it has been shown that TCR-dependent thymic
output and proliferation rates (a nonneutral BDI model) influence the
measured clone count patterns \cite{Dessalles2019}. These processes
form and maintain a diverse T cell receptor repertoire, which is
usually characterized by its richness. Unlike the barcode abundances
\deleted{in} arising during hematopoiesis, the neutral BDI processes are not
able to capture the shapes of the measured TCR clone counts.
%
%
It is also known that T cell residence times depend on interactions
between tissues and T cell receptors. Thus, different clones of T
cells are expected to be differentially spatially distributed in the
body. Hence, diversity metrics should be defined within and between
\emph{habitats}, much like that in ecology.

Finally, it is known that T cell richness decreases with age
\cite{YATES2012,YATES2018,Lewkiewicz2018,AGING2018}. Qualitatively, a
loss of diversity has been predicted within the multispecies BDI
process by assuming a decreasing thymic output rate with age. Even
when the thymus is \replaced{abruptly}{completely} shut down, the
diversity of the T cell repertoire slowly decreases as successive
clones go extinct and the clone abundance distribution \added{slowly}
\emph{coarsens}.  In humans, since the overall T cell population is
primarily maintained by proliferation rather than thymic
immigration~\cite{denBraber2012}, the reduction in diversity is
fortunately a slow process.
\begin{figure*}[htb]
\centering
\includegraphics[width=4.9in]{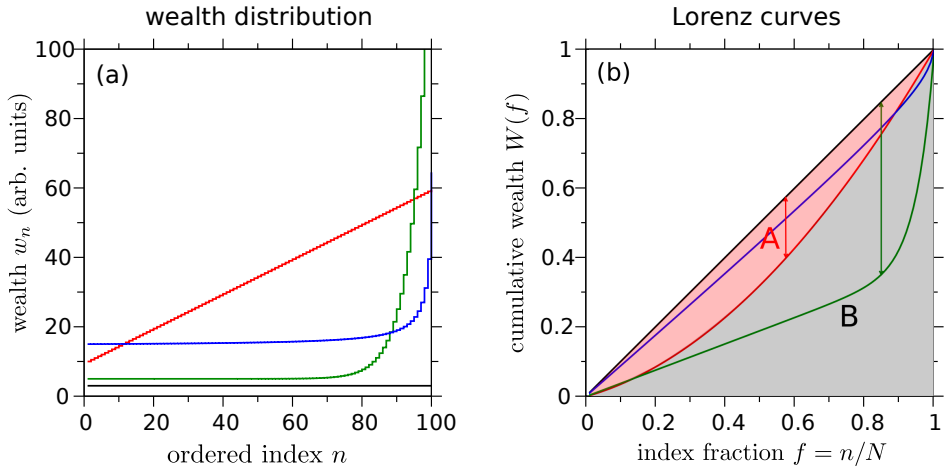}
\caption{(a) Ordering of all $N=100$ individuals in increasing wealth
  or income. The hypothetical wealth distributions plotted are
  $w_{i}=3$ (equal wealth, black curve), $w_{i} = 10+ (i-1)/2$ (linear
  distribution, red), $w_{i} = 5+e^{i/5 - 15}-e^{-14.8}$ (green), and
  $w_{i} = 14.5 + 50/(101-i)$ (blue). The latter three represent
  distributions with some amount of inequity. (b) These inequalities
  can be visually quantified by their corresponding Lorenz curves,
  plotted as the relative fraction of the population $f$. The Lorenz
  curve for a perfectly uniform wealth distribution is given by the
  straight diagonal line. The area between the diagonal equality line
  and any other Lorenz curve can be used to visualize the Gini
  coefficient of the associated wealth distribution. The Gini
  coefficient, $\mbox{Gini}=A/(A+B)$, is calculated by dividing the
difference in areas between the equality line and the Lorenz curve in
question ($A$) by the total area ($A+B=1/2$) under the equality
curves.  The `Robin Hood' index is defined as the maximum difference
between the equality line and a given Lorenz curve, and is indicated
by arrow for the red and green Lorenz curves.
\label{FIG_WEALTH}}
\end{figure*}
\subsection{Societal Applications of Diversity: Wealth distributions}
\label{sec:societal}
Metrics associated with diversity have been naturally applied in human
social
contexts~\cite{neckerman2007inequality,bailey1990social,balch2000hierarchic,ENI},
including physical, cultural,
educational~\cite{BELLCURVE,domina2017categorical}, and economic
settings. For example, the distribution of wealth is the chief metric
in many economic and political studies. As with all applications, data
collection, sampling, and delineating differences in attributes are
main research challenges.

Wealth \replaced{and}{or} income, unlike species, are essentially
continuous and ordered quantities, and can be described by many
indices designed by economists to measure different wealth attributes
of a population. Distinct from cellular or ecological contexts,
socio-economic diversity is also often discussed in terms of
`inequality,' `evenness,' or `polarization.'  Diversity or
`inequality' indices in the socioeconomic setting usually invoke a
number of additional assumptions\added{:}

\begin{itemize}
\item Individual identities are irrelevant:  This is analogous to
  barcoding studies of a singular cell type in which the barcode
  identity is not important.

\item Size and total wealth invariance: The diversity is invariant to
  the total population size. Only proportions of the total population
  that are associated with a proportion of the total wealth are
  relevant.

\item Dalton principle: Any inequality index should increase if any
  amount of wealth is transferred from an entity to one with higher
  existing wealth.
\end{itemize}

Mathematically, one starts by ordering the wealth or income of a
  population of $N$ entities $w_{1} \leq w_{2} \ldots \leq w_{i} \leq
  w_{i+1}, \ldots \leq w_{N}$. For large $N$, the rescaled wealth
  distribution $w(f)\equiv w_{fN}$ is a function of the relative
  fraction of the total population $f=n/N \in[0,1]$. Furthermore, we
  can define a normalized wealth distribution or density

\begin{equation}
\tilde{w}(f)  = {w(f) \over W_{\rm T}}, \quad W_{\rm T} =\sum_{i=1}^{N}w_{i}\approx
\int_{0}^{1} w(f')\dd f',
\end{equation}
and the corresponding cumulative distribution
\begin{equation}
W_i=\frac{1}{W_T}\sum_{j=1}^i w_j
\end{equation}
or
\begin{equation}
W(f) = \int_{0}^{f}\tilde{w}(f')\dd f' \equiv 
\frac{1}{W_{\rm T}}\int_{0}^{f}w(f')\dd f'.
\end{equation}
The functions $W(f)$ are known as `Lorenz-consistent' if they
satisfy the above assumptions~\cite{LORENZ1905}. Four representative
Lorenz consistent raw wealth distributions are shown in
Fig.~\ref{FIG_WEALTH}(a) as functions of the individual index. In
Fig.~\ref{FIG_WEALTH}(b), we plot the continuous cumulative rescaled
wealth distribution $W(f)$ as a function of the relative population
fraction $f$ corresponding to the wealth distributions shown in
Fig.~\ref{FIG_WEALTH}(a).  From any ordered distribution, we can
define a so-called `Lorenz curve' that illustrates many indices
graphically. The Lorenz curve is defined as the cumulative wealth of
all individuals of a \emph{relative} index $f=n/N$ and lower.

Many indices can be visualized by the Lorenz curves. For example, the
Gini index~\cite{gini1912variabilita,GASTWIRTH1972} for the red
distribution (linear wealth) in Fig.~\ref{FIG_WEALTH}(a) is calculated
by the area of the red shaded region ($A$) divided by the area under
the equality curve ($A+B=1/2$): $\mbox{Gini}=A/(A+B) = 2A$. In a
society where every person receives the same income, the Gini index
equals zero. However, if the total wealth is concentrated in only one
out of $N$ entities, $\mbox{Gini}=1-2/N$. This motivates one to define the
Gini index for discrete cumulative wealth values $W_i$ according to

\begin{equation}
\mbox{Gini} = 1-{2 \over N}\sum_{i=1}^{N}W_i,
\end{equation}
while the `Hoover' or `Robin Hood' index defined by
\cite{HOOVER1936,atkinson1992economic,kennedy1996income}
\begin{equation}
H = \max_{f} \left\{\vert f - W(f)\vert\right\}
\end{equation}
is the Legendre transform at $f^{\ast}$, the fraction of individuals
corresponding to $\dd W(f)/\dd f\vert_{f=f^{\ast}} = 1$. For the two
Lorenz curves in Fig.~\ref{FIG_WEALTH}(b), the Robin Hood index is
indicated by the two corresponding arrows.

The Robin Hood index happens to be a specific case of the
Kolmogorov-Smirnoff statistic as defined in Eq.~(\ref{KSMETRIC}) for
two cumulative distributions. For convex functions $W(f)$
\replaced{that satisfy}{such that} $W(0)=0$, $W(1)=1$, the index $H$
corresponds to the fraction of the total wealth that needs to be
distributed in order to achieve uniform wealth. This can be seen by
considering the wealth $w_i$ up to an index $n^\ast$ such that
$w_i\leq N^{-1}$ for all $i\leq n^\ast$. The total wealth that needs
to be redistributed to obtain equal wealth fractions $N^{-1}$ for
every individual is
\begin{eqnarray}
H=\sum_{i=1}^{n^\ast} \left({1 \over N}-w_i\right) & 
= \frac{n^\ast}{N}-W_{n^\ast} \nonumber \\
\: & \approx f^\ast-W(f^\ast).
\end{eqnarray}
Another possibility is to sum over all entities $w_i$ according to
\begin{eqnarray}
H & = {1\over 2}\sum_{i=1}^{N}\left\vert {1\over N}-w_{i}\right\vert 
\nonumber \\
\: & \approx  {1\over 2}\int_{0}^{1}\left\vert 1-w(f)\right\vert \dd f \nonumber \\
&=\frac{1}{2}\left[\int_{0}^{f^{\ast}} (1- w(f)) \, \dd f + \int_{f^{\ast}}^1 (w(f)-1) \, \dd f
\right] \nonumber \\
& =f^{\ast} - W(f^\ast).
\end{eqnarray}
The specific, local redistribution is not specified but it would be
intriguing to cast it in the language of optimal transport and
Wasserstein distances \cite{OPTIMAL1}. This way, one might also define
costs to wealth redistribution.

It is also possible to quantify inequity according to the Theil
index~\cite{theil1972statistical,novotny2007measurement,lasarte2014decomposition}
\begin{equation}
  T = \frac{1}{N} \sum_{i=1}^N \frac{w_i}{\mathds{E}[w]}
  \log \left(\frac{w_i}{\mathds{E}[w]}\right),
\end{equation}
which corresponds to a relative entropy as defined in
Eq.~(\ref{KLDIVERGENCE}). In this case, the entropy of the
distribution of $w_i$ is measured with respect to the
\replaced{expected}{expectation} value
$\mathds{E}[w]=N^{-1}\sum_{i=1}^N w_i$. If $\sum_{i=1}^N w_i = 1$, we
may interpret $w_i$ as the probability of finding an individual in
income class $i$, and $\mathds{E}[w]=N^{-1}$ corresponds to the
relative share of equally distributed wealth. Naturally, many
\replaced{other}{others} measures for inequality have been defined
\replaced{by}{my} numerous authors \replaced{focusing}{focussing} on
specific socioeconomic areas \cite{DEMAIO}.

However, typical inequality indices do not convey any judgment, belief
system, or behavioral propensity on measured inequity and thus may not
capture typical social concepts. In an effort to better quantify
concepts such as inequity or
`polarization,'~\cite{bottcher2019competing} a sociologists have
proposed a number of \textit{polarization indices} that are argued to
be more directly correlated with social tension and unrest. For
example, Esteban and Ray~\cite{ESTEBAN_RAY1994,DUCLOS2004} developed a
measure of polarization to account for clusters within which
individuals are more similar in an attribute $x$ (such as wealth)
\replaced{than}{that} they are between clusters.  While there may be
many ways to define polarization, imposing a few reasonable features
and constraints can narrow down the allowable forms. First, they
assume an `identity-alienation framework' in which an individual
also identifies with his own distribution $f(x)$ at value $x$. An
effective `antagonism' of an individual with attribute $x$ towards
those with attribute $y$ is defined as $T[f(x), d]$ where a simple
form for the distance is $d = \vert x-y\vert$. The polarization $P$ is
then assumed to take the form
\begin{equation}
P[f] =\int \int T[f(x),\vert x-y\vert]f(x)f(y) \, \dd x \, \dd y.
\end{equation}
By imposing axioms that the polarization (i) cannot increase if the
distribution is \emph{squeezed} (compressed towards its peak), (ii) must
increase if two non-overlapping distributions are moved farther apart,
and (iii) the polarization should be invariant to scalings of the
total population. Using these constraints, the polarization can be
more explicitly defined as
\begin{equation}
P[f] = \int \int f^{1+\alpha}(x)f(y)\vert x-y\vert \, \dd x \, \dd y,
\label{PF_ALPHA}
\end{equation}
where $1/4 \leq \alpha \leq 1$ \cite{DUCLOS2004} (Esteban and Ray
\cite{ESTEBAN_RAY1994} and Kawada, Nakamura, and Sunada \cite{KAWADA}
\replaced{found}{find} $0\leq \alpha < 1.6$ using slightly different
assumptions). The parameter $\alpha$ describes the amount of
`polarization sensitivity.' It measures identification of a
population with its distribution and distinguishes polarization from
other standard inequity measures such as the Gini index (when $\alpha
= 0$ \cite{ESTEBAN_RAY1994}) or Simpson's index.  Also, note that when
$\alpha=0$, the form of $P[f]$ resembles the total potential energy of
a system of particles \replaced{that are}{which is} distributed
according to $f(x)$ and exhibits an interaction energy $\vert
x-y\vert$. The discrete analogue of Eq.~(\ref{PF_ALPHA}) is $P[f]
\propto \sum_{i,j}f_{i}^{1+\alpha}f_{j}\vert x_{i}-x_{j}\vert$, for
which the individuals $i,j$ can be generalized to groups. In empirical
studies, the Esteban and Ray polarization measure is given by
\begin{equation}
P_{\rm{ER}}[f]\propto \sum_{i,j} \pi_i^{1+\alpha} \pi_j \vert \mu_i - \mu_j\vert,
\label{ESTEBANRAY}
\end{equation}
where
\begin{eqnarray}
\pi_i &= \int_{x_{i-1}}^{x_i} \!\!\! f(x) \, \dd x\quad \rm{and} \quad
\mu_i &= \frac{1}{\pi_i} \int_{x_{i-1}}^{x_i} \!\!\!x f(x) \, \dd x,
\end{eqnarray}
are the relative frequency and the mean of the wealth in group
  $i$, respectively~\cite{DAMBROSIO_WOLFF}. D'Ambrosio and Wolff
  suggested replacing the difference of mean wealths in
  Eq.~(\ref{ESTEBANRAY}) by the Kolmogorov measure of variation
  distance~\cite{DAMBROSIO_WOLFF,d2001household}
\begin{equation}
\mathrm{Kov}_{i j} = \frac{1}{2} \int \vert f_i(y)-f_j(y) \vert \, \dd y,
\end{equation}
to obtain
\begin{equation}
P_{\rm{DW}}[f]\propto \sum_{i,j} \pi_i^{1+\alpha} \pi_j \rm{Kov}_{ij}.
\end{equation}
Additional indices have been proposed, including a class of
polarizations by Tsui and Wang~\cite{WANGTSUI} of the form
\begin{equation}
P_{\rm TW}({\bf x}) = {1\over N}\sum_{i=1}^{N}\psi(d_{i}), 
\quad d_{i} = \bigg| {x_{i} - m(\x) \over m(\x)}\bigg|,
\end{equation}
where $\psi$ is a smooth function of the rescaled distance $d_{i}$.
The median income $m(\x)$ is computed from the individual incomes
$x_i$ ($1\leq i \leq N$).

Many of these polarization metrics can in fact be expressed in terms of the 
Gini coefficient. For example, the Foster--Wolfson polarization index 
is defined as~\cite{WOLFSON}
\begin{equation}
P_{\rm FW}({\bf x}) = ({\rm Gini}_B-{\rm Gini}_W)(\mu({\bf x})/m({\bf x})),
\end{equation}
where $\mu({\bf x})$ is the corresponding mean income, and the
subscript indices $B$ and $W$ denote the between and within group Gini
coefficients. According to the definition of $P_{\rm FW}({\bf x})$,
inequity differs from polarization in the following way: The Gini
index as the sum of ${\rm Gini}_B$ and ${\rm Gini}_W$ quantifies the
unequal distribution of wealth in a society whereas polarization is
measured in terms of the difference of ${\rm Gini}_B$ and ${\rm
  Gini}_W$. Thus, an increase in within-group inequality leads to a
larger total inequality, but \deleted{to} a lower polarization.  A more refined
understanding of socioeconomic diversity will need to consider
multiple classes of attributes, including possible geographic or
spatial distributions. 


The described polarization measures are relevant not only in the
context of wealth distributions, but they are also able to provide
important insights \replaced{to}{in} other sociological phenomena
associated with the notion of diversity. As one example, quantitative
measures of polarization are applicable to examine factors that
influence the cohesiveness of groups~\cite{mas2013short}. \deleted{In
  addition, diversity measures may help to identify mechanisms which
  lead to inequality among different social groups in our education
  system~\cite{domina2017categorical}.} In this context, the social
entropy theory aims to quantitatively compare diversity across social
systems such as societies, organizations, and individual
groups~\cite{bailey1990social,bailey1990social2,balch2000hierarchic}.

\section{Summary and Discussion}
\label{sec:summary}
%
%

  \begin{table*}[htp]
      \caption{Summary of fundamental, commonly used diversity
        measures. The variable $q$ indicates the order of the
        corresponding Hill number $^{q}\! D$ as defined in
        Eq.~(\ref{DEQN}). }
      \label{tab:summary}
      \vspace{1cm}
\begin{adjustbox}{angle=90}
  \begin{tabular}{|L{0.19\textwidth}| L{0.21\textwidth}| L{0.22\textwidth} | L{0.23\textwidth}| L{0.25\textwidth}|}
\hline
\textbf{Measure}  &\textbf{Interpretation} & \textbf{Application}  & \textbf{Advantages}  & \textbf{Disadvantages}  \\ \hline
Species number $n_i$ 
& number of entities of type $i$
& evolutionary and population models 
& straightforward interpretation in models
& keeping track of species identity may be unrealistic
\\ \hline
Species abundance (clone count) $c_k$ 
& number of species of size $k$
& models of self-assembly/nucleation~\cite{WATTIS1998,LAKATOS_JCP,LEIQI}; 
characterization of population in barcoding studies~\cite{goyal2015mechanisms}
& directly related to richness, useful when clone identity is not
important & no clone identity information, insensitive to exchange of
populations between clones \\ \hline
Richness ($^{0}\! D$)  & total number of distinguishable species &
conservation planning;
assessment of ecosystems~\cite{purvis2000getting} & straightforward
mathematical definition and interpretation & 
maximally affected by
small sampling; all species are treated equally \\ \hline
Evenness ($^{1}\! D$) & uniformity of relative abundances of species
in a population & characterization of ecosystems and inequity in
societies; Theil
index~\cite{purvis2000getting,theil1972statistical,novotny2007measurement,lasarte2014decomposition}
& straightforward mathematical definition and interpretation; similar
to entropy & affected by sampling \\ \hline
Simpson's diversity ($^{2}\! D$) & probability that two randomly drawn
entities are of the same species
& 
characterization of cell populations~\cite{rosenfeld2018computational,venturi2007methods} & 
less affected by sampling
& more intricate mathematical definition \& less interpretability
\\  \hline
$^{q}\! D$,  $q > 2$ & N/A & characterization
  of more frequent species in a population; Berger-Parker index~\cite{berger1970diversity} & significantly less affected by sampling
& no intuitive interpretation 
%
%
  \\ \hline
  Lorenz curve & cumulative relative wealth & economics, wealth distributions
  & fundamental mathematical object & no identity information (like
  ordered clone counts)
  \\ \hline
  Gini index & deviation of Lorenz curves from absolute equality & 
  population-level wealth inequality & easily understood  & no identity information,
  values are subjectively interpreted
  \\ \hline
  Hoover/Robin Hood index & KS statistic between Lorenz curve and
  equality line & population-level wealth inequality & easily
  understood & no identity information, values are subjectively
  interpreted
  \\ \hline
  \end{tabular}
\end{adjustbox}
  \end{table*}


Quantifying the diversity of a given population in terms of a single
measure such as richness does not fully describe the underlying
distribution of species or other properties. Various diversity
measures have been developed and tailored to specific applications in
different fields including ecology, biology, and
economics. Mathematically, one can describe populations in terms of
species numbers $n_i$ (number of entities of type $i$) or clone counts
$c_k$ (number of species of size $k$).  Hill numbers $^{q}\!D$ provide
a framework to unify some common diversity indices that are based on a
species-number description. Hill numbers with large values of $q$ put
more weight on common species whereas small values of $q$ yield
measures that are more sensitive to rarer species. This implies that
measures such as richness ($q=0$) and evenness ($q=1$) are more prone
to sampling effects than Simpson's diversity index ($q=2$) or Hill
numbers with $q>2$~\cite{soetaertl1990sample}.  In
Table~\ref{tab:summary}, we summarize some common diversity measures,
their applications, and advantages and disadvantages.


In conclusion, we have provided an overview of the most relevant
measures of diversity and their information-theoretic counterparts.
We then summarized common applications of diversity indices in
biological and ecological systems. Despite the ambiguity in the
definitions and the variety of \deleted{different} diversity measures
\cite{simpson1949measurement,hurlbert1971nonconcept,sepkoski1988alpha,purvis2000getting,WHITTAKER2001,ricotta2005through,sarkar2006ecological},
the concept is \added{still} of great importance for the monitoring
of ecosystems and in the context of conservation planning
\cite{heywood1995global,courtillot1996effects,john1998rates,sala2000global,margules2000systematic,alroy2004sepkoski,sarkar2006ecological}.

We also described the importance of a quantitative treatment of
diversity for experiments in the study of the gut microbiome, stem
cell barcoding, and the adaptive immune system. Finally, we discussed
examples of the application of diversity measures in human social
systems including the characterization of wealth distributions in
societies and measures of political or cultural polarization.
Scientific conclusion\added{s} in these fields\deleted{,} and in
ecology\deleted{,} are particularly sensitive to sampling and
measurements.
However, accurate measurements \cite{MORTON_VS_GOULD}, meaningful
classification, spatial resolution \cite{GOLDENFELD2006}, and
informative sampling protocols \cite{WILLIS2015,LENNON2016} remain
elusive across almost all fields.  Sometimes, as
\replaced{illustrated}{shown for example} in Fig.~\ref{CK_TCELL}(b),
different measures even lead to contradictory conclusions
\cite{nagendra2002opposite}.  There is no golden rule in choosing a
unique metric for a specific situation, as the sampling effects also
depend on the underlying unknown clone-count
distribution~\cite{soetaertl1990sample}.  It is recommended that one
\replaced{cross-checks}{considers} different metrics \deleted{and
  cross-checks their values} while bearing in mind how sampling
effects may impact diversity measures differently.

\section{Acknowledgments}
This work was supported in part by grants from the NSF through grant
DMS-1814364, the Army Research Office through grant W911NF-18-1-0345,
and the National Institutes of Health through grant R01HL146552. The
authors also thank Greg Huber for insightful discussions.

\section*{References}

\bibliographystyle{iopart-num} 
\bibliography{references3}


\end{document}